\documentclass[10pt,final]{IEEEtran}

\usepackage[dvips]{graphicx}
\usepackage{float}
\usepackage{times}
\usepackage{amsmath}
\usepackage{amsfonts}
\usepackage{balance}
\usepackage{color}
\usepackage[normalem]{ulem}
\usepackage{cite}
\renewcommand{\baselinestretch}{1.0}

\begin{document}
%
\title{Location and Orientation Optimisation for Spatially Stretched Tripole Arrays Based on Compressive Sensing}

\author{\IEEEauthorblockN{Matthew Hawes$^{a}$, Lyudmila Mihaylova$^{a}$ and Wei Liu$^{b}$}\\[0.3cm]
\IEEEauthorblockA{$^{a}$ Department of
Automatic Control and Systems Engineering, University of Sheffield, S1 3JD, UK\\$^{b}$ Department of Electronic and Electrical Engineering, University of Sheffield, S1 3JD, UK\\
{\{m.hawes, l.s.mihaylova, w.liu\}@sheffield.ac.uk }}

}

\maketitle

\markboth{If citing this work please cite the version published by IEEE Transactions on Signal Processing, DOI: 10.1109/TSP.2017.2655479}%
{Hawes \MakeLowercase{\textit{et al.}}: SSSTA}

\begin{abstract}

The design of sparse spatially stretched tripole arrays is an important but also challenging task
and this paper proposes for the very
first time efficient solutions to this problem. Unlike for the
design of traditional sparse antenna arrays, the developed approaches optimise both the dipole locations
and orientations. The novelty of the paper consists in formulating these optimisation problems into a form that
can be solved by the proposed compressive sensing and Bayesian compressive sensing based approaches. The performance of the developed
approaches is validated and it is shown that accurate approximation of a reference response can be
achieved with a 67$\%$ reduction in the number of dipoles required as compared to an equivalent uniform
spatially stretched tripole array, leading to a significant reduction in the cost associated with the resulting arrays.

\end{abstract}
\section{Introduction}
\subsection{Related Work}
For uniform linear arrays (ULAs), an adjacent antenna separation of no larger than half of the
operating wavelength is used to avoid the introduction of grating lobes \cite{vantrees02a,liu2010}.
 This can become prohibitive in terms of the cost associated with the number of antennas required.
  Instead, sparse arrays become a desirable alternative due to the fact that the nonuniform nature of their
  adjacent antenna separations avoids grating lobes even when the mean adjacent antenna separation is greater
  than half the operating wavelength \cite{Jarske88}.

However, the sidelobe behaviour of sparse arrays is unpredictable.  This means that optimisation of the
antenna locations is required in order to achieve a desired beam response.  Such optimisation can be achieved by stochastic optimisation methods such as
genetic algorithms (GAs)
\cite{Haupt94,Yan97a,Hawes12}, and simulated
annealing (SA)\cite{Trucco99a,Repetto06}.  Difference sets and almost difference
sets have also been successfully used in the design of sparse
arrays, \cite{Oliveri09,Oliveri10}, and merged with GAs to help give
an improved performance, \cite{Caorsi04,Oliveri11}.  The
disadvantage of GAs, and similar design methods,  is the potentially
long computation time and the possibility of convergence to a
non-optimal solution.

More recently, the area of compressive sensing (CS) has been
explored \cite{Candes06}, and CS-based methods have been proposed in
the design of traditional sparse arrays
\cite{Prisco11,Carin09,Cen10b,Hawes14c,Hawes14b,Hawes15b}. CS theory says that when certain conditions are met it is possible to recover
some signals from fewer measurements than used by traditional
methods \cite{Candes06}. It is possible to use CS to design sparse sensor arrays by obtaining a close approximation of a desired beam response using as few array elements as possible.

Further work has also shown that it is possible to improve the
sparseness of a solution by considering a reweighted $l_{1}$ norm
minimisation problem \cite{Hawes14c,Candes08,Fuchs12,Prisco12}.  The aim of
these methods is to bring the minimisation of the $l_{1}$ norm of
the weight coefficients closer to that of the minimisation of the
$l_{0}$ norm.  To do this an iterative method is required to solve a
series of reweighted $l_{1}$ minimisation problems, where locations
with small weight coefficients are more heavily penalised than
locations with large weight coefficients.

Alternatively, the problem can be converted into a
probabilistic framework (termed Bayesian compressive sensing (BCS)) \cite{Ji08}, with some suggested advantages to BCS as compared to traditional CS based implementations.
However, an important point of interest is that the problem can be solved by the relevance vector machine
(RVM) optimisation framework \cite{Tipping01}, which is efficient to use as also supported by the
comparisons shown in the design examples section of this paper.  Additionally, using BCS can remove the
need to fine tune the error limits or sparsity associated with the implementations of CS above \cite{Carmi10}.
  Such approaches have been applied in the design of sparse
arrays with real valued and complex valued weight coefficients
\cite{Oliveri11a,Oliveri12,Viani13}, where the multi-task BCS scheme \cite{Ji09}, is applied in the case of
 complex valued weight coefficients.

The methods discussed above have been implemented assuming the arrays consist of isotropic array elements.
As a result, the polarisation of a
signal is not taken into account when considering the performance of
an array. Instead arrays based on vector sensors, \cite{Hawes15b,Zhong14}, provide a desirable alternative as they
allow the measurement of both the horizontal and vertical components of the received waveform. For example,
the vector sensors used could be crossed dipoles (two orthogonally orientated dipoles)
\cite{Hawes15b,Compton81,Li91,WangK15}, or tripoles (three orthogonally orientated dipoles) \cite{Compton81b,Liu13a}.

When tripoles are used it is possible to measure the full electromagnetic (EM) field at a given
point \cite{Liu13a}.  These arrays have been applied in the area of direction and polarisation estimation
\cite{Compton81b}.  Due to the close proximity of the three orthogonal dipoles that make up each
tripole there can be issues with mutual coupling when implemented in practice.  As a result, the concept of
spatially stretched tripoles (SST) has been developed and used in the area of direction of arrival (DOA)
estimation \cite{Liu13a}.  An SST is a tripole where the three orthogonal dipoles are spread over a
given geometry, leading to reduced mutual coupling effects.

\subsection{Contributions}

In this work for the first time the problem of designing sparse SST arrays (SSSTAs) is addressed.
Unlike for the design of traditional sparse arrays there are now two optimisation problems to solve,
i.e. finding the optimal locations and orientations for the dipoles.  It is proposed to use CS and BCS
based design methods that go beyond the state of the art in order to solve these problems.

As a result, it is now necessary to formulate the problem to include the fact that there are three potential dipoles at
each point on the sampling grid and the signal model now includes polarisation information (requiring
alterations to the CS and BCS formulations).  It is possible to avoid co-located dipoles by viewing them
as a special case of the minimum adjacent dipole separation not meeting a physical size constraint
\cite{Hawes14c}.  However, if the methods in \cite{Hawes14c} are directly applied in this case, then although
there will be a minimum spacing between antenna locations, there can still be multiple dipoles at each
location.  Therefore it is necessary to consider co-located dipoles as breaking the size constraint.
Here, the design of SSSTAs utilising the size constraint is implemented in two ways: i) An iterative minimum
distance sampling method (IMDSM) with CS and BCS; ii) an altered iterative reweighted minimisation scheme
(AIRMS).  When integrating the CS/BCS based method with the IMDSM it is also important to account for the
response due to the previously fixed dipoles when deciding what the reference response in the current
iteration is.

The remainder of this paper is structured as follows:
Section~\ref{sec:design} gives details of the proposed design methods, including the array model being used (\ref{sub:AM}), a review of CS and BCS (\ref{sub:CS} and \ref{sub:BCS}) and the proposed IMDSM and reweighted design methods for SSSTAs (\ref{sub:IMDSM} and \ref{sub:reweighted}).
In Section~\ref{sec:sim} design examples are presented to verify the effectiveness of the proposed methods and conclusions are
drawn in Section~\ref{sec:con}.

\section{Proposed Design Methods}\label{sec:design}
\subsection{Array Model}\label{sub:AM}
\begin{figure}[htb]
\begin{center}
   \includegraphics[angle=0,width=0.37\textwidth]{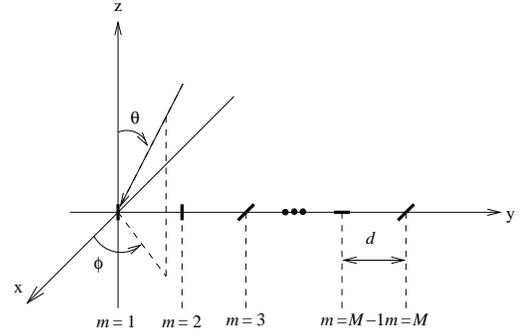}
   \caption{An example of a spatially stretched tripole array, where the short lines indicate the dipoles and the directions they are orientated in, x, y and z are the axes of a Cartesian coordinate system and $M$ is the number of dipoles used.
    \label{fig:AM}}
\end{center}
\end{figure}
Figure \ref{fig:AM} shows an example of a linear SSSTA.  $M$ possible dipole locations are spread along the y-axis with an adjacent separation of $d$.  For each possible dipole location there are three potential orientation directions, one parallel to each axis.  Also shown is a signal with its direction of
arrival (DOA) defined by the angles $\theta$ and $\phi$, with $0\leq\theta\leq\pi/2$ and $-\pi/2\leq\phi\leq\pi/2$ \cite{Compton81b,Liu13a}.  A
plane-wave signal model is assumed, i.e. the signal impinges upon
the array from the far field.

The spatial steering vector of the array is given by
\begin{eqnarray}\nonumber
\label{eq:s_s}
    \textbf{s}_{s}(\theta,\phi)&=&[1, \ldots, e^{-j2\pi d\sin\theta\sin\phi/\lambda},\\  && \ldots, e^{-j2\pi(M-1)d\sin\theta\sin\phi/\lambda}]^{T},
\end{eqnarray}
where $\lambda$ is the wavelength of interest and $\{.\}^{T}$ indicates the transpose operation.  The spatial-polarization coherent vector, which contains information about
a signal's polarisation and is given by
\cite{Compton81b,Liu13a}:
\begin{eqnarray}\label{eq:s_p}\nonumber
\textbf{s}_{p}(\theta,\phi,\gamma,\eta) &=&
\left[
  \begin{array}{c}
    \sin\gamma\cos\theta\cos\phi e^{j\eta} - \cos\gamma\sin\phi \\
    \sin\gamma\cos\theta\sin\phi e^{j\eta} - \cos\gamma\cos\phi \\
    -\sin\gamma\sin\theta e^{j\eta} \\
  \end{array}
\right] \\  &=& \left[
  \begin{array}{c}
    s_{p,x}(\theta,\phi,\gamma,\eta) \\
    s_{p,y}(\theta,\phi,\gamma,\eta) \\
    s_{p,z}(\theta,\phi,\gamma,\eta) \\
  \end{array}
\right]
 \end{eqnarray}
where $\gamma\in[0,\pi/2]$ is the auxiliary polarization angle and
$\eta\in[-\pi,\pi)$ is the polarization phase difference.

Now the array can be split into three sub-arrays, one parallel to each axis. With $f\in\{x,y,z\}$, the steering vector of each sub-array is given by:
\begin{equation}\label{eq:s_x}
\textbf{s}_{f}(\theta,\phi,\gamma,\eta) =
 s_{p,f}(\theta,\phi,\gamma,\eta))\textbf{s}_{s}(\theta,\phi).
 \end{equation}

 The response of the array is given by
\begin{equation}\label{eq:p2}
  p(\theta,\phi,\gamma,\eta)=\textbf{s}(\theta,\phi,\gamma,\eta)^{T}\textbf{w},
\end{equation}
with
\begin{eqnarray}\label{eq:w1}
  \textbf{w} &=& [w_{x,1}, w_{y,1}, w_{z,1}, \ldots, w_{x,M}, w_{y,M}, w_{z,M} ]^{H},
\end{eqnarray}
where $w_{1}=w_{x,1}$ is the complex weight coefficient for the dipole located at the point $m=1$ and orientated parallel to the $x$-axis and $\{.\}^{H}$ denotes the Hermitian transpose.  Note that for an SSSTA if $w_{x,1}\neq0$, then $w_{y,1}=w_{z,1}=0$, as there can be only one dipole present.  Similarly
\begin{eqnarray}\label{eq:s1}\nonumber
  \textbf{s}(\theta,\phi,\gamma,\eta) &=& [s_{x,1}(\theta,\phi,\gamma,\eta),s_{y,1}(\theta,\phi,\gamma,\eta),\\ \nonumber && s_{z,1}(\theta,\phi,\gamma,\eta), \ldots, s_{x,M}(\theta,\phi,\gamma,\eta),\\ &&s_{y,M}(\theta,\phi,\gamma,\eta), s_{z,M}(\theta,\phi,\gamma,\eta)]^{T},
\end{eqnarray}
where
$s_{x,1}(\theta,\phi,\gamma,\eta)$ is the contribution of the dipole located at the point $m=1$ to the overall steering vector parallel to the $x$-axis.

\subsection{Compressive Sensing for SSSTA Design}\label{sub:CS}
Suppose  $P_{r}(\theta,\phi,\gamma,\eta)$ is the desired beam response as a function of  $\theta, \phi, \gamma$ and $\eta$. Then the problem is to match the designed response to this desired response for the
full range of $\theta, \phi, \gamma$ and $\eta$ values of interest while finding the
optimised dipole locations and orientations.

First, consider Figure \ref{fig:AM} as being
a grid of potential dipole locations.  Here $M$ is a large number and sparseness is then introduced by selecting the weight
coefficients to give as few active dipoles as possible, or in other words as few non-zero valued weight coefficients as possible,
while still giving a designed response close to the desired
one.  Note, a large $M$ means it is more likely that the optimal locations will appear on the grid thus allowing for a better performance.  However, the tradeoff is that if $M$ is too large the efficiency of the algorithm deteriorates.

This problem is formulated as
\begin{eqnarray}\label{eq:min1}
    \min||\textbf{w}||_{1}\;\;\;\;\; \text{subject to}\;\;\;\;\;
    ||\textbf{p}_r-\textbf{S}\textbf{w}||_{2}\leq\alpha\;,
\end{eqnarray}
where $||\textbf{w}||_{1}$ is the $l_{1}$ norm of the weight coefficients \cite{Candes06}, $\textbf{p}_r$ is the vector holding
the desired beam response at the sampled angular and polarisation
points of interest, $\textbf{S}$ is the matrix composed of the
corresponding steering vectors, and $\alpha$ places a limit on the
allowed difference between the desired and the designed responses.  Minimising the $l_{1}$ norm has the effect of minimising the number of dipoles used, while the constraint ensures a reasonable approximation of the ideal reference response is achieved.  If the size of $\alpha$ is increased, more error can be introduced into the final response, which would be expected to allow a sparser solution to be achieved.  Note, $||.||_{2}$ indicates the $l_{2}$ norm.

In detail, $\textbf{p}_r$ and $\textbf{S}$ are respectively
given by
\begin{eqnarray}
\label{eq:pr}
    \textbf{p}_r&=&[P_r(\theta_{1},\phi_{1},\gamma_{1},\eta_{1}), \ldots, P_r(\theta_{L},\phi_{L},\gamma_{L},\eta_{L})]^{T},\\
\label{eq:S}
    \textbf{S}&=&[\textbf{s}(\theta_{1},\phi_{1},\gamma_{1},\eta_{1}),
    \ldots,
    \textbf{s}(\theta_{L},\phi_{L},\gamma_{L},\eta_{L})]^{T},
\end{eqnarray}
where $L$ is the number of points sampled at each dimension of the
desired beam response.  In this work $\textbf{p}_{r}$ is the ideal
response, i.e. a value of one for the mainlobe and zeros for the
other entries.  Note, $L$ has to be large enough to ensure all angular and polarisation points of interest are considered.

Since the coefficients are complex valued, \eqref{eq:min1} can be reformulated as a modified $l_{1}$ norm minimisation \cite{Winter05}:
\begin{eqnarray}\label{eq:mint}\nonumber
&\min& q \;\; \epsilon\;\; \mathbb{R}^{+}\\ &\text{subject to}&
    ||\textbf{p}_{r}-\textbf{S}\textbf{w}||_{2}\leq\alpha\;, \;\;\;\;|\langle\textbf{w}\rangle|_{1}\leq q
\end{eqnarray}
where
\begin{equation}\label{eq:constraint2}
    |\langle\textbf{w}\rangle|_{1}=\sum_{m=1}^{3M}||\textbf{w}_{m}||_{2}
\end{equation}
and $\textbf{w}_{m}=[R(w_{m}),I(w_{m})]^{T}$ for $m=1, \ldots 3M$ contains the real and imaginary components of the complex weight coefficient given by the $m^{th}$ entry in $\textbf{w}$.  Here, the variable $q$ has been introduced and requires minimising.  By keeping $|\langle\textbf{w}\rangle|_{1}$ less than this value the effect is to minimise the $l_{1}$ norm of all of the absolute weight coefficients.

Now decompose $q$ to $q=\sum_{m=1}^{3M}q_{m}$,
$q_{m}\in\;\;\mathbb{R}^{+}$, to reformulate \eqref{eq:mint}.  Note, the upper limit on the sum is $3M$ as there are $3$ potential dipole orientations at each location.

In vector form, $q=\textbf{1}^T\textbf{q},$ where $\textbf{1}^T=[1, \cdots, 1]$ and $\textbf{q}~=~[q_{1}, \cdots, q_{3M}]^{T}$.  Then (\ref{eq:mint}) can be rewritten as
\begin{eqnarray}\label{eq:min1t}\nonumber
&\min\limits_{\textbf{q}}& \textbf{1}^T\textbf{q}\\ \nonumber
&\text{subject to}&
    ||\textbf{p}_{r}-\textbf{S}\textbf{w}||_{2}\leq\alpha\\&&
    ||\textbf{w}_{m}||_{2}\leq q_{m},\; \;m=1, \ldots, 3M.
\end{eqnarray}
Note, a value of $q_{m}=0$, means the second constraint in \eqref{eq:min1t} ensures that the real and imaginary parts of the weight coefficient contained in $\textbf{w}_{m}$ will both be equal to zero.  This allows the desired sparsity to be introduced.

Now define
\begin{eqnarray}\label{eq:w_hat}\nonumber
  \hat{\textbf{w}} &=& [q_{1}, R(w_{x,1}), -I(w_{x,1})q_{2}, \ldots, -I(w_{z,M})]^{T},\\
\label{eq:c_hat}
    \hat{\textbf{c}} &=& [1, 0, 0, 1, 0, 0, \cdots, 1, 0, 0]^{T}\;,\\
    \label{eq:prhat}
    \hat{\textbf{p}}_{r} &=& [R(\textbf{p}_{r}),I(\textbf{p}_{r})]^{T}
\end{eqnarray}
and
\begin{equation}\label{eq:S_hat}
    \hat{\textbf{S}}=\left(
                       \begin{array}{cc}
                         \boldsymbol{0} & \boldsymbol{0}  \\
                         R(\textbf{s}_{x,1}) & I(\textbf{s}_{x,1})  \\
                         -I(\textbf{s}_{x,1}) &R(\textbf{s}_{x,1})  \\
                                                  \boldsymbol{0} & \boldsymbol{0} \\
                         R(\textbf{s}_{y,1}) & I(\textbf{s}_{y,1})  \\
                         -I(\textbf{s}_{y,1}) &R(\textbf{s}_{y,1})  \\
                         \vdots & \vdots  \\
                         R(\textbf{s}_{Z,M}) & I(\textbf{s}_{Z,M})  \\
                         -I(\textbf{s}_{Z,M}) & R(\textbf{s}_{Z,M})  \\
                       \end{array}
                     \right)^{T},
\end{equation}
where $R(.)$ is the real component and $I(.)$ is the imaginary component.  Then, the final formulation is as follows
\begin{eqnarray}\label{eq:cw}\nonumber
&\min\limits_{\hat{\textbf{w}}}& \hat{\textbf{w}}^{T}\hat{\textbf{c}}\\
\nonumber &\text{subject to}&
    ||\hat{\textbf{p}}_{r}-\hat{\textbf{S}}\hat{\textbf{w}}||_{2}\leq\alpha\\&&
    ||\textbf{w}_{m}||_{2}\leq q_{m},\; \;m=1, \cdots, 3M.
\end{eqnarray}

Note, the values $q_{m}$ for $m=1, \ldots, 3M$ are included with the weight coefficients
in $\hat{\textbf{w}}$.  This is so that it is not necessary to predefine their values, instead the
algorithm finds them at the same time as the optimised weight coefficients.  As a result, it is necessary
for the vector $\hat{\textbf{c}}$ to select the values $q_{m}$ for minimisation and the zeros are introduced
into $\hat{\textbf{S}}$ to ensure the same values do not contribute to the error between the ideal reference
response and the achieved response in the first constraint in \eqref{eq:cw}.  Finally, as the weight
coefficients have been split into real and imaginary parts, the response given by the product
$\hat{\textbf{S}}\hat{\textbf{w}}^{H}$ will contain the real and imaginary parts of the achieved response
separately.  This means the reference pattern has to be split in a similar manner giving \eqref{eq:prhat}.

However, unlike the $l_{0}$ norm, the
$l_{1}$ norm does not penalise all non-zero valued coefficients
equally.  Instead, larger coefficients are penalised more heavily.
To further improve the sparseness of the array and get a better
approximation of the $l_{0}$ norm minimisation, large reweighting terms can be applied to the smaller weight coefficients so that they are penalised more heavily \cite{Hawes14c,Hawes14b,Candes08,Fuchs12,Prisco12}.

When applied to the above modified $l_{1}$ norm minimisation problem we get the following
\begin{eqnarray}\label{eq:recw}\nonumber
&\min\limits_{\hat{\textbf{w}}}& \hat{\textbf{w}}^{T}\hat{\textbf{c}}\\
\nonumber &\text{subject to}&
    ||\hat{\textbf{p}}_{r}-\hat{\textbf{S}}\hat{\textbf{w}}||_{2}\leq\alpha\\&&
    \delta_{m}^{i}||\textbf{w}_{m}||_{2}\leq q_{m},\; \;m=1, \cdots, 3M,
\end{eqnarray}
where now $\hat{\textbf{c}} = [\delta^{i}_{1}, 0, 0, \delta^{i}_{2}, 0, 0, \ldots, \delta^{i}_{3M},0, 0]^{T}$ and $\delta^{i}_{m}~=~(|w_{m}^{i-1}|+\epsilon)^{-1}.$  Here $i$ is the current iteration, $\hat{\textbf{w}}$ holds the current estimate of the
weight coefficients, $w_{m}^{i-1}$ contains the weight coefficients, from the previous iteration, for the $m^{th}$ dipole
and $\epsilon$ is a small value roughly equal to the minimum desired
weight coefficient. The iterative algorithm would then follow the
steps below:
\begin{enumerate}
  \item Set $i = 0$ and find an initial estimate of the weight coefficients
  by solving (\ref{eq:cw}).
  \item $i = i + 1$, and find the reweighting terms $\delta_{m}^{i}$.
  \item Solve (\ref{eq:recw}).
  \item Repeat steps 2 to 3 until
  $||\textbf{w}^{i}||_{0}=||\textbf{w}^{i-1}||_{0}=||\textbf{w}^{i-2}||_{0}$ i.e. until the
  number of active locations has remained the same for three
  iterations.  Here define $\textbf{w}^{i}=[w_{1}^{i}, w_{2}^{i}, \ldots, w_{3M}^{i}]^{T}$.
\end{enumerate}
The addition of the reweighting term, which is calculated using
coefficients from the previous iteration, means all non-zero valued
coefficients are penalised in a more uniform manner.

It is worth noting that as it stands the solutions to (\ref{eq:cw}) and (\ref{eq:recw}) do not strictly give an SSSTA in the result.  This is because currently there is no way of guaranteeing there can only be a single dipole at a given location.  In other words the proposed methods are in effect finding a sparse weight coefficient vector without considering the locations of the associated dipoles.  The methods detailed in Section \ref{sub:IMDSM} and Section \ref{sub:reweighted} can both be used to overcome this issue and ensure that there are no co-located dipoles, guaranteeing an SSSTA.

\subsection{Bayesian Compressive Sensing for SSSTA Design}\label{sub:BCS}
When considering BCS for sparse array design, \cite{Oliveri11a,Oliveri12,Viani13,Oliveri12a}, there are two formulations of BCS that can be used.  Firstly there is a single task (ST) BCS formulation \cite{Ji08} which can be implemented using a RVM \cite{Tipping01,Tipping03}.  Alternatively multi task (MT) BCS, \cite{Ji09}, can be used when there are multiple CS measurements and the statistical relationships between them can be exploited.  This could include measurements at multiple time instances, or in the case of sparse array design if multiple or complex weight coefficients have to be minimised.  As a result MT-BCS is well suited to the problem being addressed and is formulated in what follows.  However, the ST-BCS based design methodology for SSSTA design is provided in the appendix for the interested reader.

Firstly, consider matching the real and imaginary parts of the achieved array response to that of the ideal reference response:
\begin{eqnarray}\label{eq:mtbcs1}
  \hat{\textbf{p}}_{F}-\breve{\textbf{S}}\textbf{w}_{F}^{T} = \tilde{\textbf{D}}_{F},
\end{eqnarray}
where $F\in\{R,I\}$, $\tilde{\textbf{D}}_{R}$ and $\tilde{\textbf{D}}_{I}$ are zero mean Gaussian error vectors, with a variance of $\breve{\sigma}^{2}$, $\textbf{w}_{R}=R(\textbf{w})$, $\textbf{w}_{I}=-I(\textbf{w})$, $\breve{\textbf{S}}=[R(\textbf{S})^{T},I(\textbf{S})^{T}]^{T}$, $\textbf{p}_{r}=\textbf{p}_{R}+j\textbf{p}_{I}$, $\hat{\textbf{p}}_{R}=[R(\textbf{p}_{R}),I(\textbf{p}_{R})]^{T}$ and $\hat{\textbf{p}}_{I}=[R(\textbf{p}_{I}),I(\textbf{p}_{I})]^{T}$.  The problem now is to find the solutions to solve
\begin{eqnarray}\label{eq:mtbcs2}
  \textbf{w}_{F} &=&\max\limits_{\textbf{w}_{F}}\mathcal{P}(\textbf{w}_{F}|\hat{\textbf{p}}_{F}).
\end{eqnarray}

It is known that for the likelihood function $\mathcal{P}(\hat{\textbf{p}}_{F}|\textbf{w}_{F})$ and the priors $\mathcal{P}(\textbf{w}_{F})$ and $\mathcal{P}(\hat{\textbf{p}}_{F})$, the following applies
\begin{equation}\label{eq:mtbcs3}
  \mathcal{P}(\textbf{w}_{F}|\hat{\textbf{p}}_{F})=\frac{\mathcal{P}(\hat{\textbf{p}}_{F}|\textbf{w}_{F})\mathcal{P}(\textbf{w}_{F})}{\mathcal{P}(\hat{\textbf{p}}_{F})}.
\end{equation}
This allows the problem to be written as
\begin{equation}\label{eq:mtbcs4}
  \textbf{w}_{F} =\max\limits_{\textbf{w}_{F}}\mathcal{P}\Bigg(\frac{\mathcal{P}(\hat{\textbf{p}}_{F}|\textbf{w}_{F})\mathcal{P}(\textbf{w}_{F})}{\mathcal{P}(\hat{\textbf{p}}_{F})} \Bigg).
\end{equation}

The prior $\mathcal{P}(\textbf{w}_{R})$ is the same as $\mathcal{P}(\textbf{w}_{I})$ to model the relationship between the real and imaginary parts of the weight coefficients, while still enforcing sparsity.  It is given by $\mathcal{P}(\textbf{w}_{F})$ and found as follows:
\begin{equation}\label{eq:mtbcs5}
  \mathcal{P}(\textbf{w}_{F})=\int\mathcal{P}(\textbf{w}_{F}|\breve{\textbf{a}},\breve{\sigma}^{2})\mathcal{P}(\breve{\textbf{a}})\mathcal{P}(\breve{\sigma}^{2})d\breve{\textbf{a}}d\breve{\sigma}^{2},
\end{equation}
where $\mathcal{P}(\breve{\textbf{a}})$ is the multi-task shared hyperpriors, $\breve{\textbf{a}}~=~[\breve{a}_{1}, \breve{a}_{1}, ..., \breve{a}_{1}]^{T}$, given by a Gamma distribution, and $\mathcal{P}(\breve{\sigma}^{2})$ is a shared Gamma hierarchial prior, where
\begin{equation}\label{eq:mtbcs6}
  \mathcal{P}(\textbf{w}_{F}|\breve{\textbf{a}},\breve{\sigma}^{2})=(2\pi\breve{\sigma})^{-3M}\prod_{m=1}^{3M}\sqrt{\breve{a}_{m}}e^{-\frac{\breve{a}_{m}w_{F,m}^{2}}{2\breve{\sigma}^{2}}},
\end{equation}
gives
\begin{eqnarray}\label{eq:mtbcs7}\nonumber
  \textbf{w}_{F,opt} =\;\;\;\;\;\;\;\;\;\;\;\;\;\;\;\;\;\;\;\;\;\;\;\;\;\;\;\;\;\;\;\;\;\;\;\\  \max\limits_{\textbf{w}_{F}}\bigg(\int\frac{\mathcal{P}(\textbf{w}_{F}|\breve{\textbf{a}},\breve{\sigma}^{2})\mathcal{P}(\hat{\textbf{p}}_{F}|\textbf{w}_{F})\mathcal{P}(\breve{\textbf{a}})\mathcal{P}(\breve{\sigma}^{2})}{\mathcal{P}(\hat{\textbf{p}}_{F})}d\breve{\textbf{a}}d\breve{\sigma}^{2} \Bigg),
\end{eqnarray}
which after integrating over $\breve{\sigma}^{2}$ and simplifying gives:
\begin{equation}\label{eq:mtbcs8}
  \textbf{w}_{F,opt} = \max\limits_{\textbf{w}_{F}}\bigg(\int\mathcal{P}(\textbf{w}_{F}|\hat{\textbf{p}}_{F},\breve{\textbf{a}})\mathcal{P}(\breve{\textbf{a}}|\hat{\textbf{p}}_{F})d\breve{\textbf{a}} \Bigg).
\end{equation}
Equation \eqref{eq:mtbcs6} considers $3M$ points as there are three potential dipoles at each location.

Note,
\begin{equation}\label{eq:mtbcs9}
  \mathcal{P}(\textbf{w}_{F}|\hat{\textbf{p}}_{F},\breve{\textbf{a}})=\int\mathcal{P}(\textbf{w}_{F}|\hat{\textbf{p}}_{F},\breve{\textbf{a}},\breve{\sigma}^{2})\mathcal{P}(\breve{\sigma}^{2})d\breve{\sigma}^{2}
\end{equation}
and from Bayes' theorem
\begin{eqnarray}\label{eq:mtbcs10}\nonumber
 &\;& \mathcal{P}(\textbf{w}_{F}|\hat{\textbf{p}}_{F},\breve{\textbf{a}},\breve{\sigma}^{2})\mathcal{P}(\breve{\sigma}^{2})=\\ &&\;\;\;\;\;\;\;\;\;\frac{\mathcal{P}(\hat{\textbf{p}}_{F}|\textbf{w}_{F},\breve{\sigma}^{2})\mathcal{P}(\textbf{w}_{F}|\breve{\textbf{a}},\breve{\sigma}^{2})\mathcal{P}(\breve{\sigma}^{2})}{\int\mathcal{P}(\hat{\textbf{p}}_{F}|\textbf{w}_{F},\breve{\sigma}^{2})\mathcal{P}(\textbf{w}_{F}|\breve{\textbf{a}},\breve{\sigma}^{2})d\textbf{w}_{F}}.
\end{eqnarray}

From (\ref{eq:mtbcs6}), the fact that a Gamma hierarchial prior is placed on $\mathcal{P}(\breve{\sigma}^{2})$ and the fact that $\mathcal{P}(\textbf{w}_{F}|\hat{\textbf{p}}_{F},\breve{\sigma}^{2})$ can be modelled as a Gaussian likelihood, then
\small
\begin{eqnarray}\label{eq:mtbcs11} \nonumber
  \mathcal{P}(\textbf{w}_{F}|\hat{\textbf{p}}_{F},\breve{\textbf{a}}) = \bigg(\int_{0}^{\infty}t^{\beta_{MT-1}+(3M/2)-1}e^{-t}dt\bigg)\times\\ \nonumber
   \frac{\big(1+\frac{1}{2\beta_{MT-2}}(\textbf{w}_{F}-\hat{\boldsymbol\mu}_{F})^{T}\hat{\boldsymbol\Sigma}^{-1}(\textbf{w}_{F}-\hat{\boldsymbol\mu}_{F})\big)^{-(\beta_{MT-2}+(3M/2))}}{\big( \int_{0}^{\infty}t^{\beta_{MT-1}-1}e^{-t}dt \big)\big(2\pi\beta_{MT-2}\big)^{(3M/2)}\sqrt{\det(\hat{\boldsymbol\Sigma})}},\\
\end{eqnarray}
\normalsize
where $\beta_{MT-1}$ and $\beta_{MT-2}$ are parameters associated with the MT-BCS process chosen to encourage sparsity.  In \eqref{eq:mtbcs11} the mean and covariance are given by:
\begin{eqnarray}\label{eq:mtbcs12}
  \hat{\boldsymbol\mu}_{F}&=&\hat{\boldsymbol\Sigma}\breve{\textbf{S}}^{T}\hat{\textbf{p}}_{F}, \\ \label{eq:mtbcs13}
  \hat{\boldsymbol\Sigma}&=&\bigg(\hat{\textbf{A}}+\breve{\textbf{S}}^{T}\breve{\textbf{S}}\bigg)^{-1},
\end{eqnarray}
respectively, where $\hat{\textbf{A}}=\text{diag}(\breve{\textbf{a}})=\text{diag}(\breve{a}_{1}, \breve{a}_{2}, \ldots, \breve{a}_{3M})$.  Note, this gives a Student's t-distribution for $\mathcal{P}(\textbf{w}_{F}|\hat{\textbf{p}}_{F},\breve{\textbf{a}})$.

When considering the remaining term in (\ref{eq:mtbcs8}) a delta function approximation can be used \cite{Oliveri12}.  This is because a closed-form solution is not possible.  Note,
\begin{eqnarray}\label{eq:mtbcs14}\nonumber
  \mathcal{P}(\breve{\textbf{a}}|\hat{\textbf{p}}_{F})&\propto&\mathcal{P}(\hat{\textbf{p}}_{F}|\breve{\textbf{a}})\mathcal{P}(\breve{\textbf{a}}) \\ \nonumber
  &\propto&\bigg(\int\mathcal{P}(\hat{\textbf{p}}_{F}|\textbf{w}_{F},\breve{\sigma}^{2})\mathcal{P}(\textbf{w}_{F}|\breve{\textbf{a}},\breve{\sigma}^{2})\\ &&\;\;\;\;\;\;\;\;\;\times\mathcal{P}(\breve{\sigma}^{2})d\textbf{w}_{F}d\breve{\sigma}^{2} \bigg)\mathcal{P}(\breve{\textbf{a}}),
\end{eqnarray}
with a mode given by
\begin{equation}\label{eq:mtbcs15}
  \breve{\textbf{a}}_{opt}=\max\limits_{\breve{\textbf{a}}}\mathcal{L}(\breve{\textbf{a}}),
\end{equation}
where\small
\begin{eqnarray}\label{eq:mtbcs16}\nonumber
  \mathcal{L}(\breve{\textbf{a}})=-\frac{1}{2}\sum\limits_{F}\Bigg(\log\bigg(\big|\textbf{I}+\breve{\textbf{S}}\hat{\textbf{A}}^{-1}\breve{\textbf{S}}^{T}\big| \bigg)+(3M+2\beta_{MT-1})\\ \times\log\bigg(\hat{\textbf{p}}^{T}_{F}\big(\textbf{I}+\breve{\textbf{S}}\hat{\textbf{A}}^{-1}\breve{\textbf{S}}^{T} \big)\hat{\textbf{p}}_{F}+2\beta_{MT-2} \bigg) \Bigg).
\end{eqnarray}\normalsize
As the mode of a student-t distribution is equal to its mean the resulting weight coefficients are given by \cite{Oliveri12}
\begin{equation}\label{eq:mtbcs17}
  \textbf{w}_{F,opt} = \bigg(\text{diag}(\breve{\textbf{a}}_{opt})+\breve{\textbf{S}}^{T}\breve{\textbf{S}} \bigg)^{-1}\breve{\textbf{S}}^{T}\hat{\textbf{p}}_{F}.
\end{equation}
The final optimal weight coefficient vector is then given by
\begin{equation}\label{eq:mtbcs}
  \textbf{w}_{MT,opt} = \textbf{w}_{R,opt}+j\textbf{w}_{I,opt}.
\end{equation}

Note, that as for the CS formulation discussed in the previous subsection the MT-BCS scheme detailed here is unable to guarantee an SSSTA as an outcome.  This is because it is in effect finding a sparse weight coefficient vector without considering where the associated dipoles are located.  As a result, it is possible that there could be multiple dipoles present at the optimised locations (optimised locations refers to the locations with one or more non-zero valued weight coefficients).  This means the desired reduction in mutual coupling effects when implemented in practice will not be achieved.  Instead to ensure an SSSTA the methods discussed in the following subsections should be considered.

\subsection{Iterative Minimum Distance Sampling Method for SSSTAs}\label{sub:IMDSM}
In the above two formulations, there is no way to ensure that an SSSTA is achieved.  This is due to the fact that only the weight coefficients associated with a given dipole are minimised, rather than considering if there are any co-located dipoles.

To solve this problem it is proposed to extend the idea of imposing a physical size constraint on the optimisation from \cite{Hawes14c}.  However, when directly applied these methods only ensure that there is a minimum distance between the optimised antenna locations.  Therefore, in this instances they could not guarantee an SSSTA as there can potentially be three dipoles at each antenna location.  As a result, it is necessary to also consider the fact that co-located dipoles at a given location can also be seen as breaking the minimum separation of a physical size constraint.  In this work we use the idea of the IMDSM and AIRMS algorithms proposed in \cite{Hawes14c} to ensure an SSSTA is achieved as the final solution.

Note, that the iterative nature of the IMSDM based approaches means that the relationship between $M$ or $\alpha$ and the algorithms performance becomes less predictable.  Consider the fact that the value of $M$ used affects where the first dipole is located.  This then defines the remaining aperture, which is again sampled using $M$ grid points.  As a result the density of the sampling grid in the next iteration varies depending on where the previous dipole was placed and the value of $M$, which in turn makes it difficult to predict how the performance will be effected by $M$.  The effects of $\alpha$ can also be hard to predict for similar reasons.

\subsubsection{CS Based IMDSM}
\begin{figure}
\begin{center}
   \includegraphics[angle=0,width=0.35\textwidth]{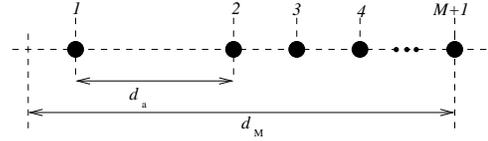}
   \caption{Illustration of the iterative sampling method used, where $d_{a}$ is the size constraint being applied, $d_{M}$ the potential aperture of the array and $\bullet$ indicates a dipole location.
    \label{fig:IMSDM}}
\end{center}
\end{figure}

To begin with, the full aperture of the array is uniformly sampled and an estimate of the weight coefficients found using (\ref{eq:cw}), with the first cluster of dipoles that are too close together being merged to give the first location as shown in Figure \ref{fig:IMSDM}.  At this point if there are multiple dipoles at the merged location the least significant are discarded to leave a single dipole present.  The remainder of the aperture is then uniformly sampled, ensuring that the next dipole will be at least the distance of the size constraint away.  This process is then repeated until there is no room for further dipoles.

It is worth noting that this method has involved the merger of dipole locations and has the potential for some dipoles to be discarded in order to avoid co-located dipoles.  As a result the weight coefficients may no longer be optimal for the given dipole locations and orientations. However, the locations and orientations can be used to efficiently implement a fixed beamformer, by minimising the sidelobe levels while keeping a unitary response for the mainlobe location.  This is detailed below in Section \ref{subsub:redesign}.

\subsubsection{MT-BCS Based IMDSM}
In essence the same iterative procedure is followed in this instance.  The initial set of weight coefficients used to find the first cluster is instead found using the MT-BCS procedure detailed in Section \ref{sub:BCS}.  For subsequent iterations some changes have to be made to ensure that the method of solving the problem can account for the fact that some dipole locations and orientations have been fixed and will be contributing to the overall response.

As a result, consider the following
\begin{equation}\label{eq:bcsimdsm1}
  \check{\textbf{p}}_{R}= \check{\textbf{S}}\check{\textbf{w}}_{R},\;\;\;\;\;\;\;\; \check{\textbf{p}}_{I}= \check{\textbf{S}}\check{\textbf{w}}_{I},
\end{equation}
where $\check{\textbf{p}}_{R}$ and $\check{\textbf{p}}_{I}$ are found by subtracting the response due to the locations fixed in the previous iteration from the reference response in the previous iteration.  Then from the remaining uniformly sampled aperture in the current iteration we construct $\check{\textbf{S}}$ and the resulting estimate of the weight coefficients are given by $\check{\textbf{w}}=\check{\textbf{w}}_{R,opt}+j\check{\textbf{w}}_{I,opt}$.  Following the MT-BCS scheme detailed in Section \ref{sub:BCS} the solution is
\begin{eqnarray}\label{eq:bcsimdsm2}
  \check{\textbf{w}}_{F,opt} = \bigg(\text{diag}(\check{\textbf{a}}_{opt})+\check{\textbf{S}}^{T}\check{\textbf{S}} \bigg)^{-1}\check{\textbf{S}}^{T}\check{\textbf{p}}_{F}.
\end{eqnarray}

This process is repeated, with the merging and discarding of dipoles.  As a result it is again necessary to use the method for redesigning the weight coefficients detailed below.

\subsubsection{Fixed Beamformer Design for Given Dipole Locations and Orientations}\label{subsub:redesign}
After obtaining the dipole locations and orientations using the CS-IMDSM or BCS-IMDSM, it is necessary to re-design the coefficients of the array to provide a closer approximation to the desired responses.  This is a classic fixed beamformer design problem and can be solved using the method described below, which is applicable to any arbitrary array geometry.

The redesign of the weight coefficients is achieved by minimising the sidelobe levels subject to a unitary response for the mainlobe direction. This can be formulated as
\begin{eqnarray}\nonumber \label{eq:redesign1}
  &\min\limits_{\hat{\textbf{w}}_{re}}& ||\hat{\textbf{p}}_{r}-\tilde{\textbf{S}}(\hat{\textbf{w}}_{mask}\circ\hat{\textbf{w}}_{re})||_{2}\\ \nonumber
  &\text{subject to}& R(\tilde{\textbf{S}}_{ML}(\hat{\textbf{w}}_{mask}\circ\hat{\textbf{w}}_{re}))=1\\
  &&I(\tilde{\textbf{S}}_{ML}(\hat{\textbf{w}}_{mask}\circ\hat{\textbf{w}}_{re}))=0,
\end{eqnarray}
where  $\hat{\textbf{w}}_{mask}=[\textbf{w}_{mask},\textbf{w}_{mask}]^{T}$ and $\textbf{w}_{mask}$ is a series of 1s and 0s to ensure only the correct dipole orientations are used, $\hat{\textbf{w}}_{re} = [R(\textbf{w}_{re}),I(\textbf{w}_{re})]^{T},$ $\tilde{\textbf{S}} = \left(
                            \begin{array}{cc}
                              R(\textbf{S}) & -I(\textbf{S}) \\
                              I(\textbf{S}) & R(\textbf{S}) \\
                            \end{array}
                          \right)$, $\tilde{\textbf{S}}_{ML}$ only considers the mainlobe direction and $\circ$ denotes the Hadamard product.

\subsection{Altered Iterative Reweighted Minimisation Scheme for SSSTAs}\label{sub:reweighted}
To avoid the merging and discarding of dipoles as required for IMDSM, this work also proposes an AIRMS.  Here the reweighting scheme in (\ref{eq:recw}) is adapted to also penalise dipole locations that are too close together \cite{Hawes14c}.  This gives the following reweighting scheme
  \begin{eqnarray}
 \delta_{m}^{i} =
 \begin{cases}
 (|w_{m}^{i-1}|+\epsilon)^{-1} & m = 1 \\
 (|w_{m}^{i-1}|+\epsilon)^{-1} & m > 1 \text{ and constraint met} \\
 (\epsilon)^{-1} & \text{otherwise}.
 \end{cases}
  \end{eqnarray}
Now the iterative procedure is repeated until a solution that complies with the size constraint being enforced is obtained.

Unfortunately, this algorithm will not always guarantee a viable
solution, due to the presence of $\epsilon$ in the calculation of
reweighting terms. The inclusion of $\epsilon$ is required for
numerical stability, but prevents a zero weight coefficient in the
current iteration guaranteeing a zero weight coefficient in the next
iteration. Based on the authors' experience with different design
parameters, if a solution is possible it will usually be achieved in
less than $10$ iterations.

It is also hard to predict if a solution will be achieved, or the performance level achieved, based on the
selection of $M$.  This is as the choice of $M$ greatly effects how likely we are to get a solution that
meets the size constraint value.  It may be expected that increasing $M$ should allow an improvement in the
algorithms performance as it is more likely to get the optimal locations included on the sampling grid.
This also makes it more likely that two or more dipoles will be located closer together than the
size constraint making it harder to get a valid solution.

\section{Design Examples}\label{sec:sim}
This section provides design examples to verify the effectiveness of the proposed methods.  All examples are implemented on a computer with
an Intel Xeon CPU E3-1271 (3.60GHz) and 16GB of RAM.

For all of the figures that follow positive values of $\theta$
indicate the value range $\theta\in[0^\circ,\; 90^\circ]$ for
$\phi=90^{\circ}$, while negative values of $\theta\in[-90^\circ,\;
0^\circ]$ indicate an equivalent range of $\theta\in[0^\circ,\;
90^\circ]$ with $\phi=-90^{\circ}$.

  Here a broadside design example and two off-broadside design examples are considered to illustrate the effectiveness of the proposed design methods, when designing linear SSSTAs.  Although the AIRMS does not necessarily require the weight coefficients to be redesigned, they have been here in order to allow a fairer comparison between all three design methods considered.  Unless otherwise stated,
the examples consider the scenario of $M=301$ with a maximum possible aperture of 10$\lambda$.  For the design examples using MT-BCS the values of $\beta_{MT-1}$ and $\beta_{MT-2}$ are set as suggested in \cite{Ji09}, with the value of $\sigma^{2}$ being found from the CS-IMDSM and AIRMS design examples.  In this work the CS-IMDSM and AIRMS are implemented using cvx, a package for specifying and solving convex programs \cite{cvx,Grant08}.

Note, the selection of $M$ has been made to get close to the sampling density suggested in \cite{Fuchs12}, while also accounting for the fact that the proposed methods have to consider three antennas at each grid point rather than a single antenna.  As discussed for the proposed methods it is also hard to predict how changing $M$ will effect the performance of the algorithms (in the case of the AIRMS a solution is not even always guaranteed).  Experience with different design examples suggest that $M=301$ for a $10\lambda$ aperture usually ensures a suitable solution will be achieved by at least one of the three proposed methods.

For the three examples the response from an equivalent ULA is also provided as a further comparison.  To ensure
optimised dipole locations and orientations for the ULAs, solve the minimisation in \eqref{eq:redesign1}
with $\textbf{w}_{mask}=[1, 1, 1, \ldots, 1]^{T}$ to allow the three dipole orientations at each location to
be considered.  Then a new $\textbf{w}_{mask}$ is constructed in order to keep only the most significant
dipole orientations at each location.  The minimisation in \eqref{eq:redesign1} is then resolved to give the
final optimised dipole orientations and locations.

\subsection{Broadside Example}
For the broadside design example, the mainlobe is given by $\theta_{ML}=0^{\circ}$ for
$\phi_{ML}=90^{\circ}$, with the sidelobe regions defined by $\theta_{SL}=[10^{\circ},90^{\circ}]$ for
$\phi_{SL}=\pm90^{\circ}$ and being sampled every $1^{\circ}$.  The polarisation information is given by
$\gamma=45^{\circ}$ and $\eta=100^{\circ}$.  For the CS-IMDSM and AIRMS examples the value of $\alpha=0.5$
is used.
\begin{figure}
\begin{center}
   \includegraphics[angle=0,width=0.4\textwidth]{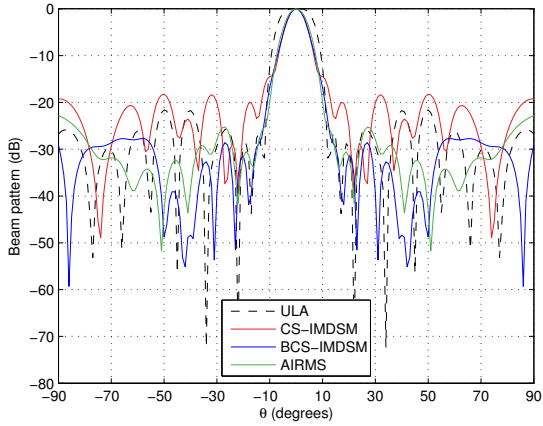}
   \caption{Designed broadside responses.
    \label{fig:broad}}
\end{center}
\end{figure}

\begin{figure}
\begin{center}
   \includegraphics[angle=0,width=0.4\textwidth]{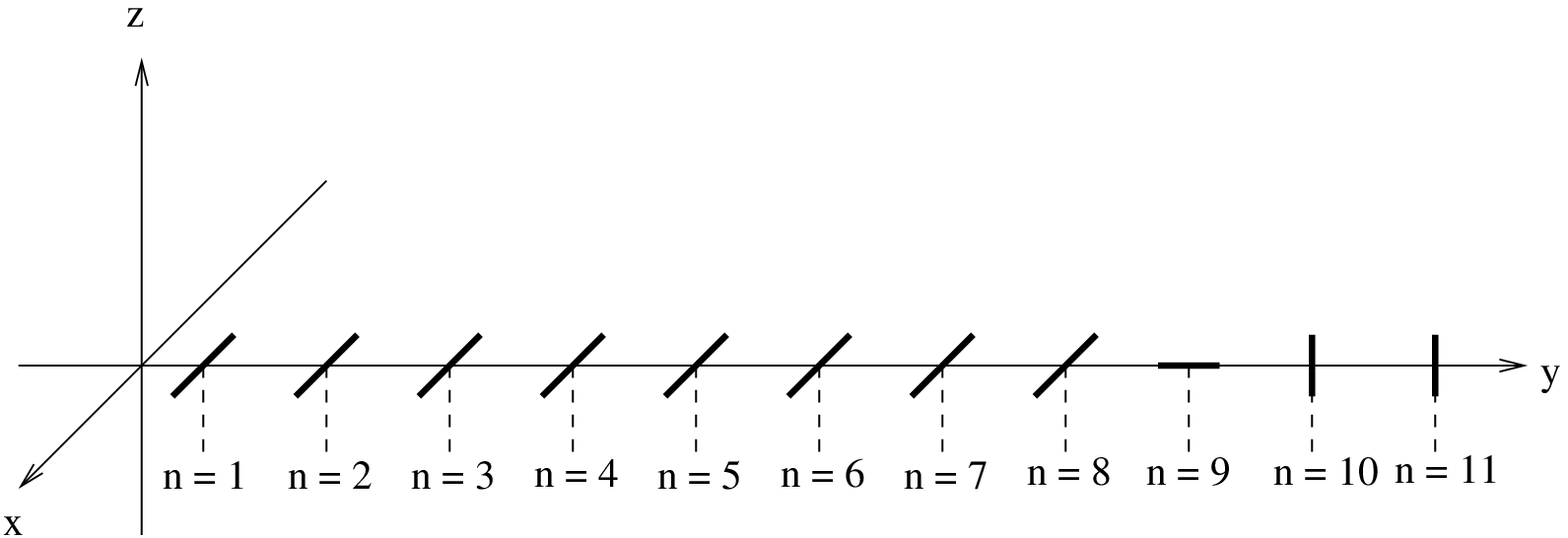}
   \caption{Dipole orientations for broadside example designed using CS-IMDSM.
    \label{fig:designed1}}
\end{center}
\end{figure}

\begin{figure}
\begin{center}
   \includegraphics[angle=0,width=0.4\textwidth]{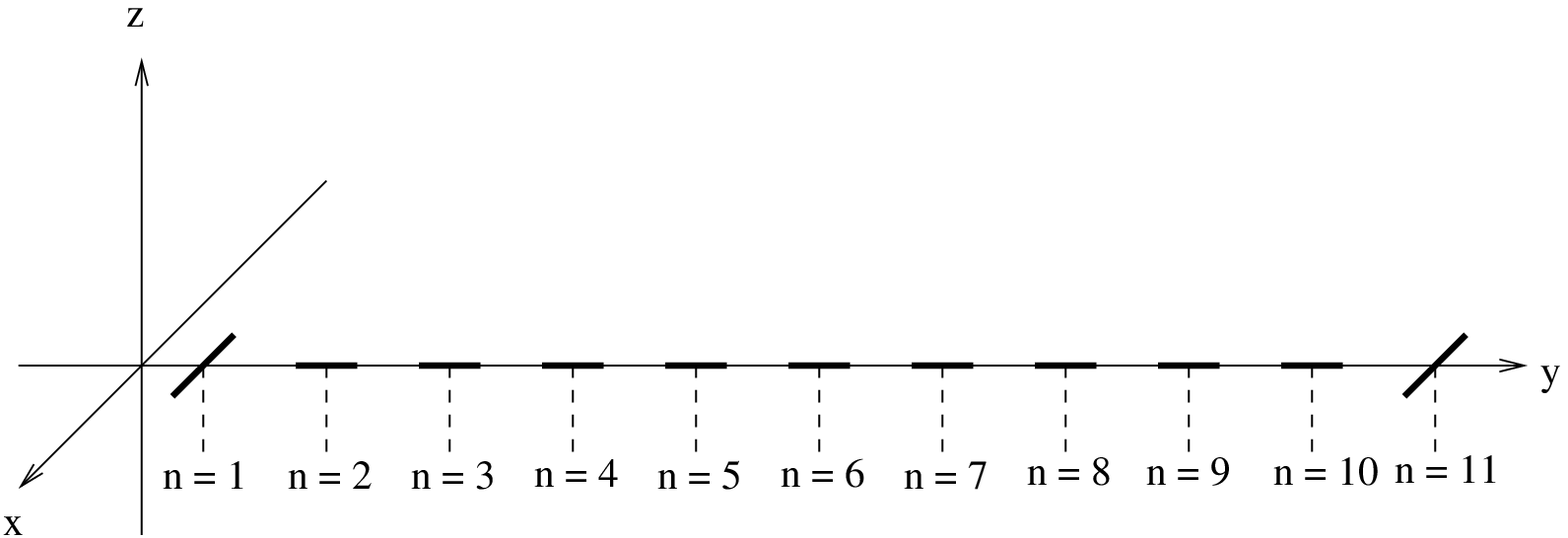}
   \caption{Dipole orientations for broadside example designed using BCS-IMDSM.
    \label{fig:designed2}}
\end{center}
\end{figure}

\begin{figure}
\begin{center}
   \includegraphics[angle=0,width=0.4\textwidth]{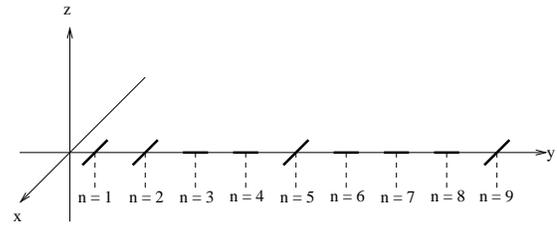}
   \caption{Dipole orientations for broadside example designed using AIRMS.
    \label{fig:designed3}}
\end{center}
\end{figure}

\begin{figure}
\begin{center}
   \includegraphics[angle=0,width=0.4\textwidth]{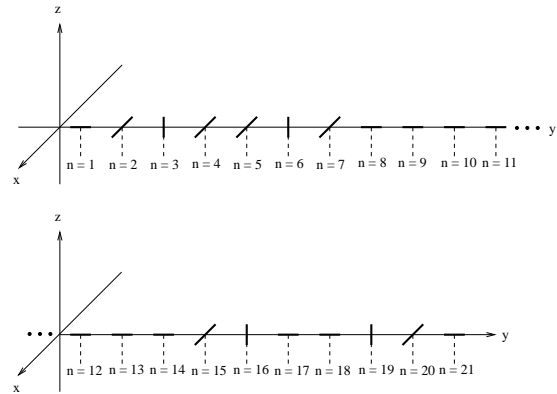}
   \caption{Dipole orientations for broadside ULA comparison example.
    \label{fig:designedula1}}
\end{center}
\end{figure}

The responses for the CS-IMDSM, BCS-IMDSM and AIRMS design examples are shown in Figure \ref{fig:broad}.  For all three of the proposed methods the correct mainlobe location has been achieved (whereas the ULA example gave a $1^{\circ}$ error), along with sufficient sidelobe attenuation.  For completeness the resulting dipole locations are shown in Tables \ref{tb:CSbroad}, \ref{tb:BCSbroad} and \ref{tb:reweightedbroad}, respectively, where it is clear the size constraint has been successfully enforced in all cases.  Figures. \ref{fig:designed1}, \ref{fig:designed2} and \ref{fig:designed3} illustrate the orientations of the dipoles for each of the three broadside examples and the ULA orientations are shown in Figure \ref{fig:designedula1}.  Note, the dipole positions shown in the figures do not accurately reflect the true dipole locations.  The true locations should instead be determined from the corresponding tables provided.

\begin{table}
\caption{\rm Dipole locations and orientations for the broadside CS-IMDSM design example.} \centering
\begin{tabular}{|c|c|c|c|c|c|c|c|}\hline
n  & $d_{n}/\lambda$ &  n & $d_{n}/\lambda$ &  n & $d_{n}/\lambda$ & n & $d_{n}/\lambda$\\
\hline
1  &  0.34  &  4 &  2.86    &  7 &  5.59  & 10 & 8.57 \\
\hline
2  & 1.18 &  5  &   3.79   &  8 &  6.53  & 11 & 9.48 \\
\hline
3  & 2.02 &   6  &   4.64  &  9 &  7.67   \\
\cline{1-6}

\end{tabular}
\label{tb:CSbroad}
\end{table}

\begin{table}
\caption{\rm Dipole locations and orientations for the broadside BCS-IMDSM design example.} \centering
\begin{tabular}{|c|c|c|c|c|c|c|c|}\hline
n  & $d_{n}/\lambda$ &  n & $d_{n}/\lambda$ &  n & $d_{n}/\lambda$ & n & $d_{n}/\lambda$\\
\hline
1  & 0.56   &  4 &   3.48   &  7 &  6.37  & 10 & 9.02 \\
\hline
2  & 1.43 &  5  &  4.48    &  8 &  7.25  & 11 & 9.89 \\
\hline
3  &2.56  &   6  &   5.44  &  9 & 8.12    \\
\cline{1-6}
\end{tabular}
\label{tb:BCSbroad}
\end{table}

\begin{table}
\caption{\rm Dipole locations and orientations for the broadside AIRMS design example.} \centering
\begin{tabular}{|c|c|c|c|c|c|c|c|}\hline
n  & $d_{n}/\lambda$ &  n & $d_{n}/\lambda$ &  n & $d_{n}/\lambda$ & n & $d_{n}/\lambda$\\
\hline
1  & 1.50   &  4 &  4.17    &  6 &  5.80  & 8 & 7.60 \\
\hline
2  & 2.30 &  5  &  5    &  7 &  6.70  & 9 &8.47  \\
\hline
3  &  3.27  \\
\cline{1-2}
\end{tabular}
\label{tb:reweightedbroad}
\end{table}

The following performance measures are summarised in Table \ref{tb:linearbroadcompare}: aperture length, mean adjacent dipole separation ($\overline{\Delta{d}}$), number of dipoles required (also given as a $\%$ reduction as compared to an equivalent ULA), $l_{2}$ norm of the error between the desired and achieved responses ($||\textbf{p}_{r}-\textbf{Sw}_{opt}||_{2}$, where $\textbf{w}_{opt}$ are the optimised weight coefficients for a given method), the amplitude of the peak sidelobe closest to the mainlobe, the computation time and the number of iterations required by each method.

Firstly, as expected, it can be seen that there are reasonably small error values, suggesting that a good match to the desired response has been achieved in each case.  For two of the three proposed methods the error between the designed and desired response is less than that for the ULA.  This suggests a better approximation of the ideal response has been achieved, despite requiring less dipoles (48$\%$ less for BCS-IMDSM and 57$\%$ less for AIRMS) and the introduction of sparsity.  It can also be seen that by comparing the values of $\overline{\Delta{d}}$ a comparable amount of sparseness has been introduced by each of the design methods, with the BCS-IMDSM performing slightly better (and also giving the lowest response error).

When considering the computation time it can be seen that there is a difference between the three methods.  The AIRMS has given a shorter computation compared to the CS-IMDSM which is explained by the fact that it requires fewer iterations as dipoles are not placed individually.  There is also a significant reduction in the computation time between the CS-IMDSM and BCS-IMDSM design examples.  This would suggest that the BCS-IMDSM design method is the more computationally efficient IMDSM based design method.  The authors' experience with different design examples also suggests that this is consistently the case and that the difference increases with the problem size.

\begin{table}
\caption{\rm Performance comparison for the broadside design examples.} \centering
\begin{tabular}{|c|c|c|c|c|}
  \hline
   & CS- & BCS- &  &\\
  Example & IMDSM & IMDSM & AIRMS &ULA\\
  \hline
  Aperture/$\lambda$ & 9.11 & 9.33 & 6.97 &10\\
  $\overline{\Delta{d}}/\lambda$ & 0.91 & 0.93 & 0.87& 0.50\\
  \hline
  Number of &  &  & & \\
  dipoles & 11 & 11 & 9&21 \\
  ($\%$ decrease) & 48  & 48 &57 &0 \\
  \hline
  Error & 1.00 & 0.43 & 0.46 &0.64\\
  \hline
  Amplitude of &  &  &  &\\
  closest sidelobe (dB)& -20.02 & -31.47 & -30.55&-26.83 \\
  \hline
  Computation &  &  &  &\\
  time (seconds) & 363.16 & 4.38 & 62.03 &1.17\\
  \hline
  Number of &  &  &  &\\
  iterations & 11 & 11 & 3 &2\\
  \hline
\end{tabular}
\label{tb:linearbroadcompare}
\end{table}

To illustrate the effects of the value of $M$ used, now consider the same design example again with the values $M=101,201$ and $M=401$, along with the original value of $M=301$.  The performance measures for the three proposed methods are summarised in Tables \ref{tb:broadCS}-\ref{tb:broadre}.

\begin{table}
\caption{\rm Performance comparison for the CS-IMDSM broadside design examples.} \centering
\begin{tabular}{|c|c|c|c|c|}
  \hline
  M & 101 & 201 & 301 & 401 \\
  \hline
  Aperture/$\lambda$ & 9.08 & 7.19 & 9.11 & 9.13\\
  $\overline{\Delta{d}}/\lambda$ & 0.91 & 0.90 & 0.91& 0.91\\
  \hline
  Number of &  &  & & \\
  dipoles & 11 & 10 & 11 &11\\
  ($\%$ decrease) & 48 & 52 & 48&48 \\
  \hline
  Error & 1.07 & 1.12 & 1.00& 1.25\\
  \hline
  Amplitude of &  &  & & \\
  closest sidelobe (dB)& -14.18 & -17.85 & -20.02 &-10.47\\
  \hline
  Computation &  &  &  &\\
  time (seconds) & 47.46 & 235.94 & 363.16 &546.89\\
  \hline
  Number of &  &  & & \\
  iterations & 11 & 10 & 11 & 11\\
  \hline
\end{tabular}
\label{tb:broadCS}
\end{table}

\begin{table}
\caption{\rm Performance comparison for the BCS-IMDSM broadside design examples.} \centering
\begin{tabular}{|c|c|c|c|c|}
  \hline
  M & 101 & 201 & 301 & 401 \\
  \hline
  Aperture/$\lambda$ & 9.05 & 9.49 & 9.33 & 8.88\\
  $\overline{\Delta{d}}/\lambda$ & 0.91 & 0.95 & 0.93& 0.89\\
  \hline
  Number of &  &  & & \\
  dipoles & 11 & 11 & 11 &11\\
  ($\%$ decrease) &48  & 48 &48 & 48\\
  \hline
  Error & 0.82 & 0.87 & 0.43& 0.81\\
  \hline
  Amplitude of &  &  & & \\
  closest sidelobe (dB)& -22.41 & -20.56 & -31.47 &-19.63\\
  \hline
  Computation &  &  &  &\\
  time (seconds) & 3.43 & 2.86 & 4.38 &38.00\\
  \hline
  Number of &  &  & & \\
  iterations & 11 & 11 & 11 & 11\\
  \hline
\end{tabular}
\label{tb:broadBCS}
\end{table}

\begin{table}
\caption{\rm Performance comparison for the AIRMS broadside design examples.} \centering
\begin{tabular}{|c|c|c|c|c|}
  \hline
  M & 101 & 201 & 301 & 401 \\
  \hline
  Aperture/$\lambda$ & NA & 6.95 & 6.97 & 6.98\\
  $\overline{\Delta{d}}/\lambda$ & NA & 0.87 & 0.87& 0.87\\
  \hline
  Number of &  &  & & \\
  dipoles & NA & 9 & 9 &9\\
  ($\%$ decrease) & NA & 57 &57 & 57\\
  \hline
  Error & NA & 0.48 & 0.46& 0.45\\
  \hline
  Amplitude of &  &  & & \\
  closest sidelobe (dB)& NA & -24.61 & -30.55 &-29.88\\
  \hline
  Computation &  &  &  &\\
  time (seconds) & NA & 34.06 & 62.03 &99.06\\
  \hline
  Number of &  &  & & \\
  iterations & NA & 2 & 3 & 2\\
  \hline
\end{tabular}
\label{tb:broadre}
\end{table}

As expected, increasing the value of $M$ has increased the computation for the three proposed design methods.
  This is because the design methods now consider a larger sampling grid for each iteration, which in turn
  means a longer computation time.  However, the effect on the other performance measures used has proven
  to be harder to predict.

For each of the design methods varying $M$ can alter the aperture of the designed array and the dipoles
required to implement it in practice.  The mean adjacent dipole separation has remained reasonably constant
and for the CS-IMSDM method the smallest separation has even occurred for the largest value of $M$. However,
for the design of traditional sparse arrays using CS-based methods, increasing the value of $M$ would lead
to an expected increase in the mean adjacent dipole separation.  This is because a denser grid will be able
to give a closer approximating to the ideal locations and as a result uses less dipole in total.  By looking
at the error between the designed responses and the ideal response, along with the amplitudes of the closest
sidelobes, it can be seen that the effect on the desirability of the designed response is similarly hard to
predict in advance.  The same is true when off-broadside examples are considered.  So for the remainder of this broadside design example and the two off-broadside design examples that follow only the original value of $M=301$ is used.

Finally, now consider the effect of $\alpha$ on the performance of the CS-IMDSM and AIRMS for the broadside design example.  Two further values of $\alpha$ will be considered, $\alpha=0.35$ and $0.65$, respectively.  The performance of the two methods for these values is summarised in Table \ref{tb:broadalpha}.  For traditional CS based problems it would be expected to see that increasing the value of $\alpha$ would increase the amount of error allowed, thus allowing extra sparsity to be introduced.  However, here we can see the iterative nature of the algorithms has made predicting the effects of $\alpha$ difficult.  As a result, in what follows a single value of $\alpha$ that gives a solution for both methods will be used in the off-broadside examples to allow a fair comparison.  Note, the reason why no results are shown for AIRMS with $\alpha=0.35$ is that no solution was obtained in this case.

\begin{table}
\caption{\rm Performance comparison for varying values of $\alpha$.} \centering
\begin{tabular}{|c|c|c|c|}
  \hline
  $\alpha$  & 0.35  & 0.65  & 0.65   \\
   (method) &  (CS-IMDSM) &  (CS-IMDSM) &  (AIRMS)  \\
  \hline
  Aperture/$\lambda$ & 5.31 & 8.89 & 6.10 \\
  $\overline{\Delta{d}}/\lambda$ & 0.88 & 0.89 & 0.87\\
  \hline
  Number of &  &  &  \\
  dipoles & 7 & 11 & 8 \\
  ($\%$ decrease) &67  & 48 & 62 \\
  \hline
  Error & 1.33 & 0.66 & 0.63\\
  \hline
  Amplitude of &  &  &  \\
  closest sidelobe (dB)& -16.83 & -21.22 & -26.26 \\
  \hline
  Computation &  &  &  \\
  time (seconds) & 379.95 & 339.11 & 71.84 \\
  \hline
  Number of &  &  &  \\
  iterations & 8 & 11 & 2 \\
  \hline
\end{tabular}
\label{tb:broadalpha}
\end{table}

\subsection{Off-Broadside Example 1}
For the first off-broadside design example consider a mainlobe location of $\theta_{ML}=60^{\circ}$ for $\phi_{ML}=90^{\circ}$, with the sidelobe regions defined as $\theta_{SL}=[0^{\circ},50^{\circ}]\bigcup[70^{\circ},90^{\circ}]$ for $\phi=90^{\circ}$ and $\theta_{SL}=[0^{\circ},90^{\circ}]$ for $\phi=-90^{\circ}$, which are sampled every $1^{\circ}$.  The polarisation information is given by $\gamma=55^{\circ}$ and $\eta=100^{\circ}$.  The value $\alpha=0.75$ is used to place a limit on the allowed error in responses.

\begin{figure}
\begin{center}
   \includegraphics[angle=0,width=0.4\textwidth]{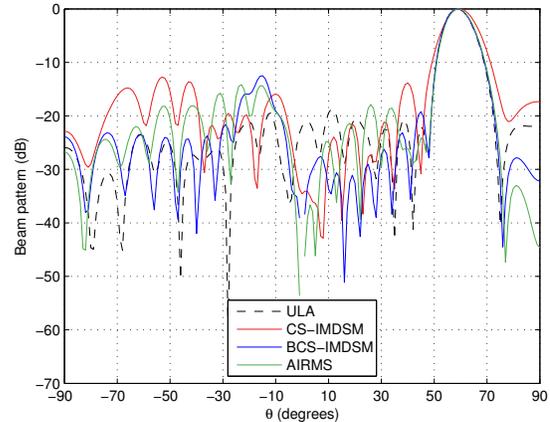}
   \caption{Designed off-broadside ($\theta_{ML}=60^{\circ}$ and $\phi_{ML}=90^{\circ}$) responses.
    \label{fig:offbroad}}
\end{center}
\end{figure}

\begin{figure}
\begin{center}
   \includegraphics[angle=0,width=0.4\textwidth]{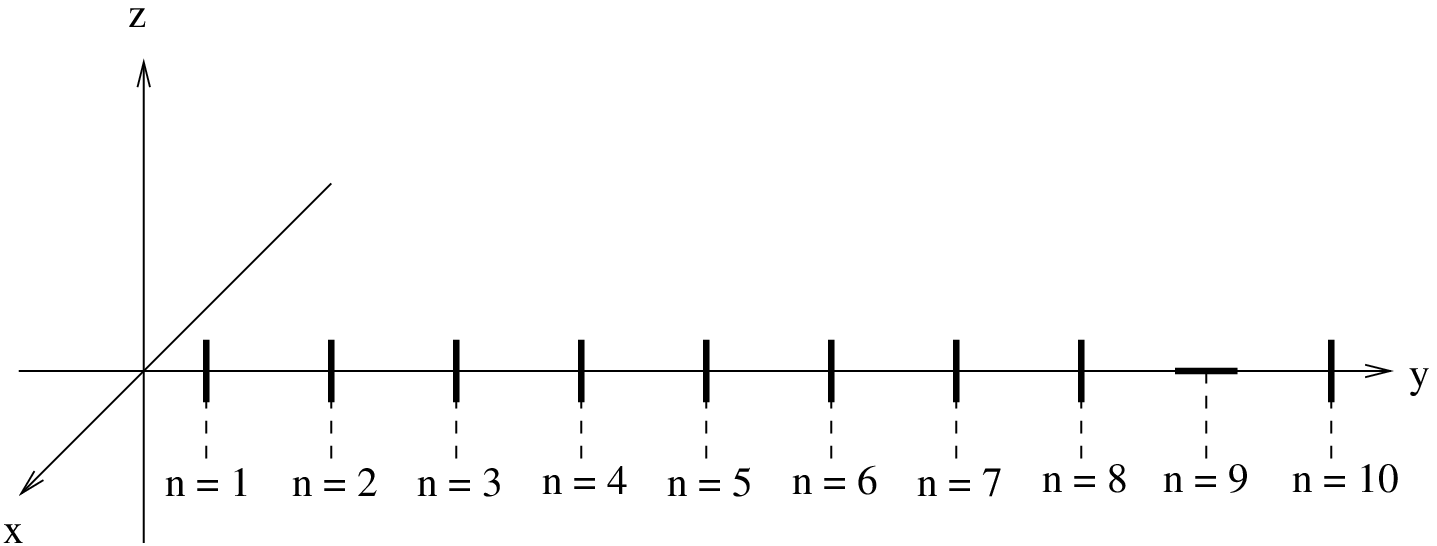}
   \caption{Dipole orientations for the off-broadside ($\theta_{ML}=60^{\circ}$ and $\phi_{ML}=90^{\circ}$) example designed using CS-IMDSM.
    \label{fig:designed4}}
\end{center}
\end{figure}

\begin{figure}
\begin{center}
   \includegraphics[angle=0,width=0.4\textwidth]{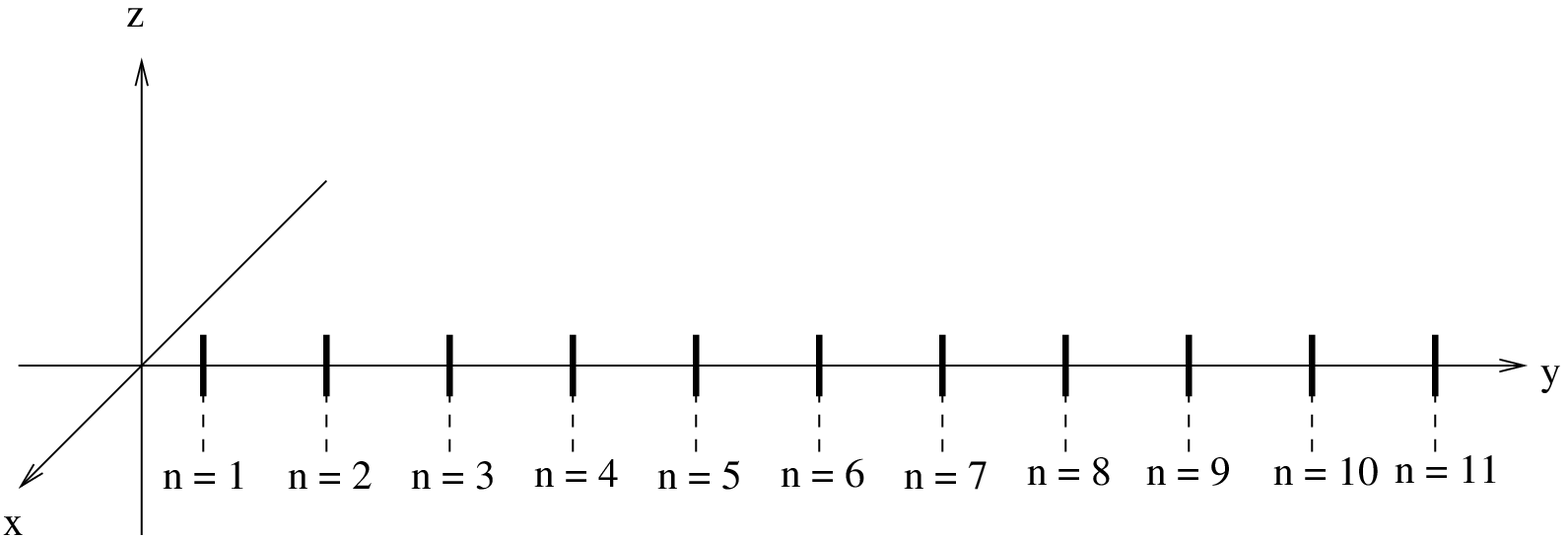}
   \caption{Dipole orientations for the off-broadside ($\theta_{ML}=60^{\circ}$ and $\phi_{ML}=90^{\circ}$) example designed using BCS-IMDSM.
    \label{fig:designed5}}
\end{center}
\end{figure}

\begin{figure}
\begin{center}
   \includegraphics[angle=0,width=0.4\textwidth]{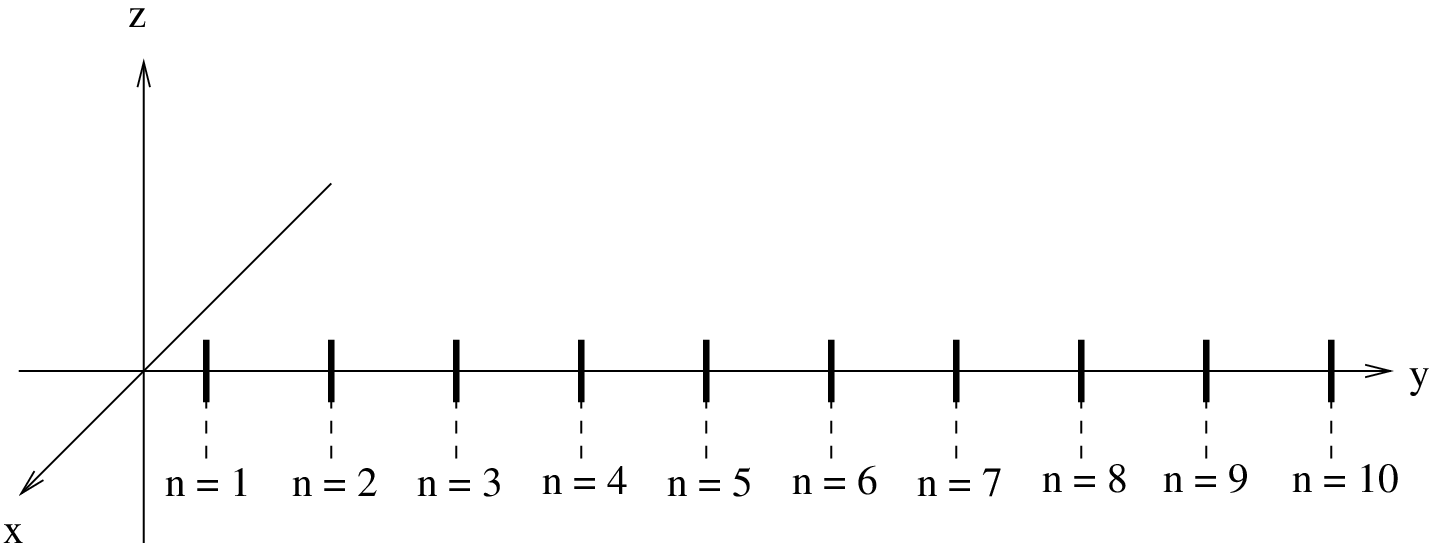}
   \caption{Dipole orientations for the off-broadside ($\theta_{ML}=60^{\circ}$ and $\phi_{ML}=90^{\circ}$) example designed using AIRMS.
    \label{fig:designed6}}
\end{center}
\end{figure}

\begin{figure}
\begin{center}
   \includegraphics[angle=0,width=0.4\textwidth]{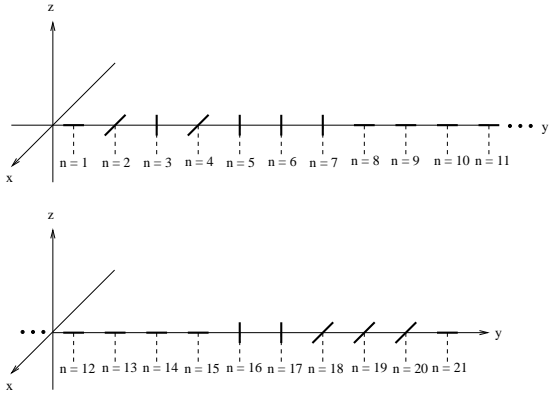}
   \caption{Dipole orientations for the ULA comparison off-broadside ($\theta_{ML}=60^{\circ}$ and $\phi_{ML}=90^{\circ}$) example.
    \label{fig:designedula2}}
\end{center}
\end{figure}

Figure \ref{fig:offbroad} shows the resulting responses for the three design examples.  The CS-IMDSM design example has the mainlobe at the correct location, while for the other two examples and the ULA comparison the mainlobe is located at $\theta=59^{\circ}$.  In all three cases sufficient sidelobe attenuation has also been achieved.  Again, for completeness the resulting dipole locations and orientations are shown in Tables \ref{tb:CSoffbroad}, \ref{tb:BCSoffbroad} and \ref{tb:reweightedoffbroad} and Figures \ref{fig:designed4}, \ref{fig:designed5} and \ref{fig:designed6}, respectively, where it is clear the size constraint has been successfully enforced in all three cases.  The comparison ULA dipole orientations are shown in Figure \ref{fig:designedula2}.  Note, the distances in the dipole orientation figures are again not intended to be accurate.  Instead, the dipole location information should be taken from the tables provided.

\begin{table}
\caption{\rm Dipole locations and orientations for the off-broadside ($\theta_{ML}=60^{\circ}$ and $\phi_{ML}=90^{\circ}$) CS-IMDSM design example.} \centering
\begin{tabular}{|c|c|c|c|c|c|c|c|}\hline
n  & $d_{n}/\lambda$ &  n & $d_{n}/\lambda$ &  n & $d_{n}/\lambda$ & n & $d_{n}/\lambda$\\
\hline
1  & 0.15   &  4 &   3.15   &  7 &  6.22  & 9 & 8.31 \\
\hline
2  &1.21  &  5  &   4.17   &  8 &  7.40  & 10 & 9.22 \\
\hline
3  & 2.22 &   6  &   5.23    \\
\cline{1-4}
\end{tabular}
\label{tb:CSoffbroad}
\end{table}

\begin{table}
\caption{\rm Dipole locations and orientations for the off-broadside ($\theta_{ML}=60^{\circ}$ and $\phi_{ML}=90^{\circ}$) BCS-IMDSM design example.} \centering
\begin{tabular}{|c|c|c|c|c|c|c|c|}\hline
n  & $d_{n}/\lambda$ &  n & $d_{n}/\lambda$ &  n & $d_{n}/\lambda$ & n & $d_{n}/\lambda$\\
\hline
1  & 0.24   &  4 &   3.22   &  7 &  5.98  & 10 & 8.55 \\
\hline
2  &1.26  &  5  &   4.16   &  8 &  6.86  & 11 & 9.37 \\
\hline
3  & 2.25 &   6  &   5.08  &  9 &   7.72  \\
\cline{1-6}
\end{tabular}
\label{tb:BCSoffbroad}
\end{table}

\begin{table}
\caption{\rm Dipole locations and orientations for the off-broadside ($\theta_{ML}=60^{\circ}$ and $\phi_{ML}=90^{\circ}$) AIRMS design example.} \centering
\begin{tabular}{|c|c|c|c|c|c|c|c|}\hline
n  & $d_{n}/\lambda$ &  n & $d_{n}/\lambda$ &  n & $d_{n}/\lambda$ & n & $d_{n}/\lambda$\\
\hline
1  & 0   &  4 &   3.37   &  7 &  6.17  & 9 & 8.27 \\
\hline
2  &1  &  5  &   4.27   &  8 &  7.20  & 10 & 9.70 \\
\hline
3  & 2.40 &   6  &   5.20    \\
\cline{1-4}
\end{tabular}
\label{tb:reweightedoffbroad}
\end{table}


\begin{table}
\caption{\rm Performance comparison for the off-broadside ($\theta_{ML}=60^{\circ}$ and $\phi_{ML}=90^{\circ}$) design examples.} \centering
\begin{tabular}{|c|c|c|c|c|}
  \hline
   & CS- & BCS- &  &\\
  Example & IMDSM & IMDSM & AIRMS &ULA\\
  \hline
  Aperture/$\lambda$ & 9.08 & 9.13 & 9.70 &10\\
  $\overline{\Delta{d}}/\lambda$ & 1.01 & 0.91 & 1.08 &0.50\\
  \hline
  Number of &  &  &  &\\
  dipoles & 10 & 11 & 12 &21\\
  ($\%$ decrease) & 52 & 48 & 62& 0\\
  \hline
  Error & 1.60 & 1.00 & 1.12 &0.89\\
  \hline
  Amplitude of &  &  &  &\\
  closest sidelobe (dB)& -13.84 & -19.20 & -24.15 &-22.02\\
  \hline
  Computation &  &  & & \\
  time (seconds) & 300.07 & 4.88 & 92.36 &1.26\\
  \hline
  Number of &  &  & & \\
  iterations & 10 & 11 & 4 &2\\
  \hline
\end{tabular}
\label{tb:linearoffbroadcompare}
\end{table}

Table \ref {tb:linearoffbroadcompare} compares the performance measures for the off-broadside design examples.  The first thing to note is that the error in the responses has significantly been increased for all three cases.  This is expected as we used a larger value of $\alpha$ and can be predicted after having looked at the three designed beam responses.  It can be seen that the BCS-IMDSM has given the most accurate estimate of the desired response (as compared to the CS-IMDSM and AIRMS), but this has come at the expense of a reduced adjacent dipole separation.  Again, the BCS-IMDSM has been shown to be the most computationally efficient of the proposed SSSTA design methods.  Finally, although the error values show a worse approximation of the ideal response has been achieved by the proposed methods, as compared to the comparison ULA, a reasonable approximation has still been achieved despite using less dipoles and the introduction of sparsity.

\begin{table}
\caption{\rm Performance comparison for CS-IMDSM with varying $\theta_{ML}$.} \centering
\begin{tabular}{|c|c|c|c|c|}
  \hline
  $\theta$ & $10^{\circ}$ & $20^{\circ}$ & $30^{\circ}$ & $40^{\circ}$ \\
  \hline
  Aperture/$\lambda$ & 9.61 & 6.92 &7.43  &7.06  \\
  $\overline{\Delta{d}}/\lambda$ & 0.80 & 1.15 & 1.06 & 1.01  \\
  \hline
  Number of &  &  &  &   \\
  dipoles & 13 & 7 &8  &8 \\
  ($\%$ decrease) & 38 & 67 & 62& 62\\
  \hline
  Error & 0.14 & 2.85 & 3.01 & 3.24  \\
  \hline
  Amplitude of &  &  &  &   \\
  closest sidelobe (dB)& -31.07 & -16.48 &  -14.96& -14.38 \\
  \hline
  Computation &  &  & &   \\
  time (seconds) &440.75  &381.60  &453.65  &370.06   \\
  \hline
  Number of &  &  & &   \\
  iterations & 13 & 8 &9  &9   \\
  \hline
  Achieved & & & & \\
   Mainlobe & $8^{\circ}$ & $20^{\circ}$ & $30^{\circ}$ &  $40^{\circ}$ \\
  \hline
  \hline
  $\theta$ & $50^{\circ}$  & $70^{\circ}$ & $80^{\circ}$ & $90^{\circ}$\\
  \hline
  Aperture/$\lambda$ &7.59 &6.04 &2.13 &2.21 \\
  $\overline{\Delta{d}}/\lambda$ &1.08 &1.01 &1.07 &1.11  \\
  \hline
  Number of  & & & &  \\
  dipoles &9 &7 & 3&  3\\
  ($\%$ decrease) & 57 &67  &86 & 86\\
  \hline
  Error & 2.25& 3.92& 4.57&5.70  \\
  \hline
  Amplitude of  & & & &  \\
  closest sidelobe (dB)&-13.77 &-14.46 &-7.24  &-4.80 \\
  \hline
  Computation  & & & &  \\
  time (seconds) & 361.64&417.89 & 215.25&330.27  \\
  \hline
  Number of  & & & &  \\
  iterations &8 & 8&4 &4  \\
  \hline
  Achieved & & & & \\
   Mainlobe & $48^{\circ}$ & $65^{\circ}$ &$90^{\circ}$  & $88^{\circ}$  \\
  \hline
\end{tabular}
\label{tb:offcs1}
\end{table}

\begin{table}
\caption{\rm Performance comparison for BCS-IMDSM with varying $\theta_{ML}$.} \centering
\begin{tabular}{|c|c|c|c|c|}
  \hline
  $\theta_{ML}$ & $10^{\circ}$ & $20^{\circ}$ & $30^{\circ}$ & $40^{\circ}$ \\
  \hline
  Aperture/$\lambda$ & 9.40 &9.36  &9.07  & 8.54 \\
  $\overline{\Delta{d}}/\lambda$ & 0.94 & 0.94 &0.91  & 0.95  \\
  \hline
  Number of &  &  &  &   \\
  dipoles &11 & 11 &11  &10 \\
  ($\%$ decrease) & 48 & 48 & 48&52 \\
  \hline
  Error &1.18  &2.04  &2.46  &2.23   \\
  \hline
  Amplitude of &  &  &  &   \\
  closest sidelobe (dB)&-10.24  &-12.23  &-12.57  &-14.66  \\
  \hline
  Computation &  &  & &   \\
  time (seconds) &5.05  &5.49  &19.06  &4.81   \\
  \hline
  Number of &  &  & &   \\
  iterations & 11 &11  &11  &11   \\
  \hline
  Achieved & & & & \\
   Mainlobe & $10^{\circ}$ &$20^{\circ}$  &$30^{\circ}$  &$40^{\circ}$   \\
  \hline
  \hline
  $\theta$ & $50^{\circ}$  & $70^{\circ}$ & $80^{\circ}$ & $90^{\circ}$\\
  \hline
  Aperture/$\lambda$ &9.40 &9.26 &9.54 &9.51 \\
  $\overline{\Delta{d}}/\lambda$ & 0.94& 0.93&0.95 &0.95  \\
  \hline
  Number of  & & & &  \\
  dipoles &11 &11 &11 &11 \\
  ($\%$ decrease) & 48 & 48 &48 &48  \\
  \hline
  Error &2.20 &2.35 &2.77 &4.47  \\
  \hline
  Amplitude of  & & & &  \\
  closest sidelobe (dB)&-10.86 &-7.32 &-7.11  &-15.58 \\
  \hline
  Computation  & & & &  \\
  time (seconds) &7.47 &5.38 &5.53 &8.06  \\
  \hline
  Number of  & & & &  \\
  iterations &11 &11 &11 &10  \\
  \hline
  Achieved & & & & \\
   Mainlobe &  $49^{\circ}$&$69^{\circ}$  &$72^{\circ}$  &$80^{\circ}$   \\
  \hline
\end{tabular}
\label{tb:offbcs1}
\end{table}

\begin{table}
\caption{\rm Performance comparison for AIRMS with varying $\theta_{ML}$.} \centering
\begin{tabular}{|c|c|c|}
  \hline
  $\theta_{ML}$ & $50^{\circ}$ & $70^{\circ}$ \\
  \hline
  Aperture/$\lambda$ &10  & 10  \\
  $\overline{\Delta{d}}/\lambda$ &1.00 & 1.00    \\
  \hline
  Number of &  &     \\
  dipoles & 11&11   \\
  ($\%$ decrease) & 48 &  48 \\
  \hline
  Error &0.97  &1.03    \\
  \hline
  Amplitude of &  &    \\
  closest sidelobe (dB)& -22.19 & -17.00  \\
  \hline
  Computation &  &     \\
  time (seconds) &65.12  &   101.58  \\
  \hline
  Number of &  &     \\
  iterations & 2 &  3   \\
  \hline
  Achieved & &  \\
   Mainlobe & $51^{\circ}$ &$69^{\circ}$     \\
  \hline
\end{tabular}
\label{tb:offairms1}
\end{table}

For completeness, now consider how the methods perform over the full range of potential off-broadside mainlobe directions.  These results are summarised in Tables \ref{tb:offcs1}-\ref{tb:offairms1}, where the designed mainlobe locations have also been added for reference.  Note, if one of the additional values of $\theta_{ML}$ is missing, it is because no solution was possible.  In addition, the comparison ULA results are summarised in Table \ref{tb:offula1}.

Here, it can be seen that there are varying performance levels for the three methods, helping illustrate that the same method is not always guaranteed to perform the best.  The first thing that can be seen is that the AIRMS has not managed to get a solution for the majority of the values of $\theta_{ML}$. However, the important thing is that one of the three solutions always appears to give an acceptable approximation of the reference response, with a reduction in the number of dipoles required, for each angle of interest.  Also, if desired a uniform SST array can be designed using the method details provided for the ULA comparisons.  One thing that does seem constant is that the BCS-IMDSM is the most efficient of the three proposed methods.  As similar patterns can be expected for the second off-broadside design example, only one value of $\theta_{ML}$ will be considered.

\begin{table}
\caption{\rm Performance comparison for ULA with varying $\theta_{ML}$.} \centering
\begin{tabular}{|c|c|c|c|c|}
  \hline
  $\theta_{ML}$ & $10^{\circ}$ & $20^{\circ}$ & $30^{\circ}$ & $40^{\circ}$ \\
  \hline
  Error & 0.52 &0.43  &0.59  & 0.90 \\
  \hline
  Amplitude of &  &  &  &   \\
  closest sidelobe (dB)&-25.71  &-22.51  &-22.40  &-26.95  \\
  \hline
  Computation &  &  & &   \\
  time (seconds) &3.56  &1.75  &1.72  &1.75   \\
  \hline
  Number of &  &  & &   \\
  iterations & 2 & 2 & 2 &  2 \\
  \hline
  Achieved & & & & \\
   Mainlobe & $11^{\circ}$ &$20^{\circ}$  &32$^{\circ}$  &38$^{\circ}$   \\
  \hline
  \hline
  $\theta$ & $50^{\circ}$  & $70^{\circ}$ & $80^{\circ}$ & $90^{\circ}$\\
  \hline
  Error &0.86 & 1.41& 1.70& 1.77 \\
  \hline
  Amplitude of  & & & &  \\
  closest sidelobe (dB)&-21.55 &-15.61 &-18.68  & -14.84\\
  \hline
  Computation  & & & &  \\
  time (seconds) &1.76 &1.80 & 2.11&1.73  \\
  \hline
  Number of  & & & &  \\
  iterations &2 &2 &2 &2  \\
  \hline
  Achieved & & & & \\
   Mainlobe &  $50^{\circ}$&$67^{\circ}$  &$84^{\circ}$  &$90^{\circ}$   \\
  \hline
\end{tabular}
\label{tb:offula1}
\end{table}

\subsection{Off-broadside Example 2}
\begin{figure}
\begin{center}
   \includegraphics[angle=0,width=0.4\textwidth]{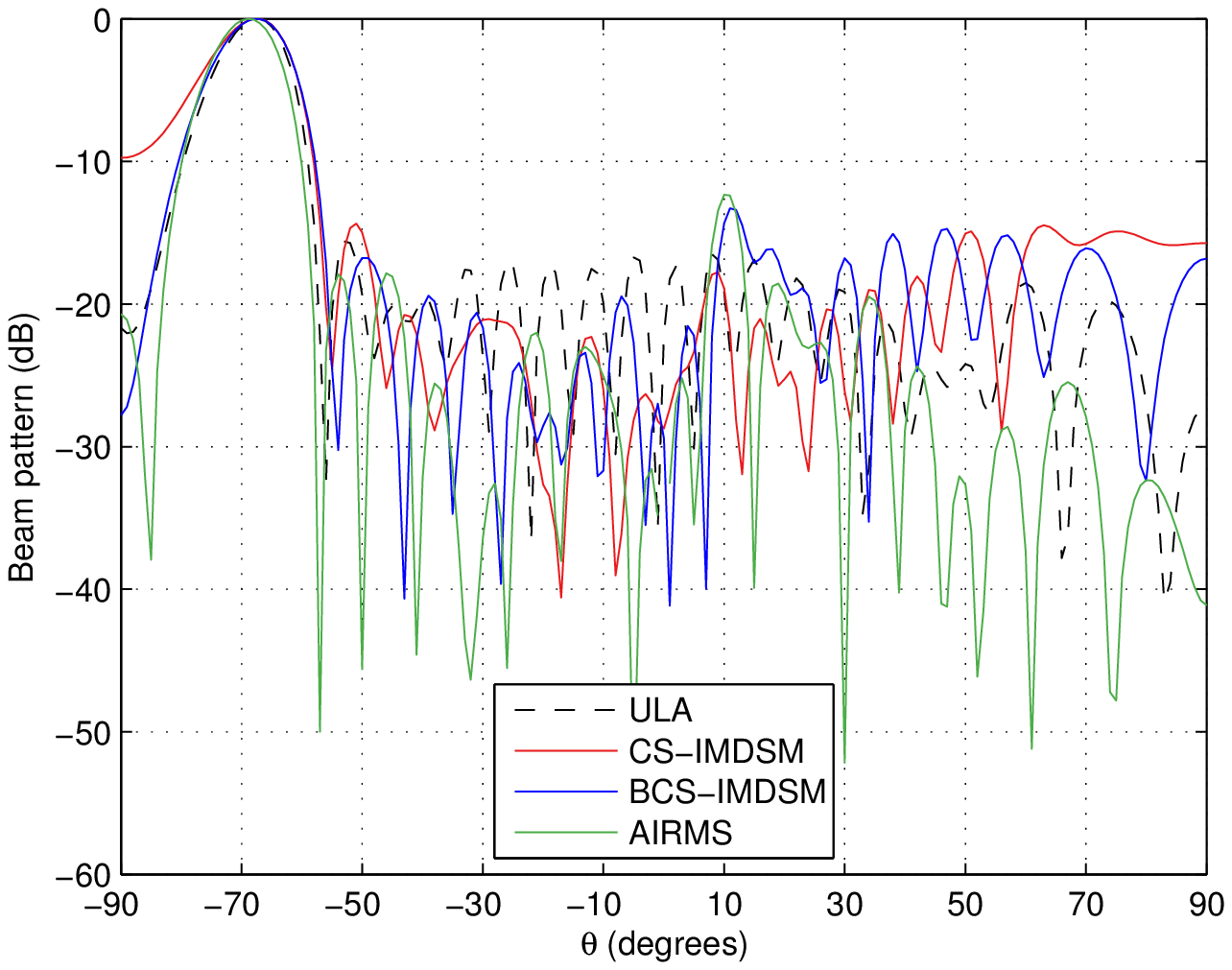}
   \caption{Designed off-broadside ($\theta_{ML}=70^{\circ}$ and $\phi_{ML}=-90^{\circ}$) responses.
    \label{fig:offbroad2}}
\end{center}
\end{figure}

In the third design example, the mainlobe is defined by $\theta_{ML}=70^{\circ}$ and $\phi_{ML}=-90^{\circ}$, with the sidelobe regions given by $\theta_{SL}=[0^{\circ},90^{\circ}]$ for $\phi_{SL}=90^{\circ}$ and $\theta_{SL}=[0^{\circ},60^{\circ}]\bigcup[80^{\circ},90^{\circ}]$ for $\phi_{SL}=-90^{\circ}$.  The value of $\theta$ is then sampled every $1^{\circ}$ in the sidelobe regions.  Finally, consider the values $\gamma=60^{\circ}, \eta=-10^{\circ}$ and $\alpha=0.8$.  This results in the responses shown in Figure \ref{fig:offbroad2}, with the various performance measures being summarised in Table \ref{tb:linearoffbroadcompare2}.

\begin{table}
\caption{\rm Performance comparison for the off-broadside ($\theta_{ML}=70^{\circ}$ and $\phi_{ML}=-90^{\circ}$) design examples.} \centering
\begin{tabular}{|c|c|c|c|c|}
  \hline
   & CS- & BCS- & & \\
  Example & IMDSM & IMDSM & AIRMS & ULA\\
  \hline
  Aperture/$\lambda$ & 9.69 & 9.30 & 10 & 10\\
  $\overline{\Delta{d}}/\lambda$ & 0.97 & 0.93 & 0.91 & 0.50\\
  \hline
  Number of &  &  &  &\\
  dipoles & 11 & 11 & 12 &21\\
  ($\%$ decrease) & 48 & 48 &43 & 0\\
  \hline
  Error & 2.08 & 1.71 & 1.03 &1.42\\
  \hline
  Amplitude of &  &  &  &\\
  closest sidelobe (dB)& -14.36 & -16.78 & -17.92 &-15.73\\
  \hline
  Computation &  &  &  &\\
  time (seconds) & 374.07 & 3.75 & 67.35 &1.24\\
  \hline
  Number of &  &  &  &\\
  iterations & 11 & 11 & 4 &2\\
  \hline
\end{tabular}
\label{tb:linearoffbroadcompare2}
\end{table}

\begin{figure}
\begin{center}
   \includegraphics[angle=0,width=0.4\textwidth]{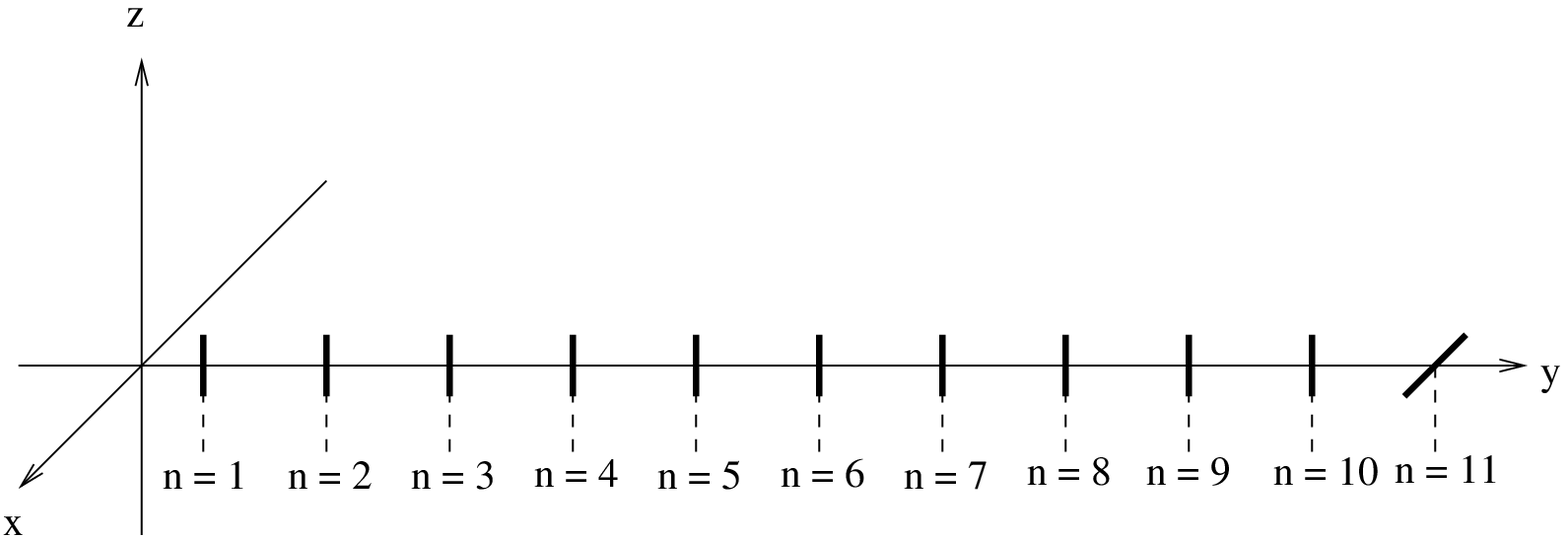}
   \caption{Dipole orientations for the off-broadside ($\theta_{ML}=70^{\circ}$ and $\phi_{ML}=-90^{\circ}$) example designed using CS-IMDSM.
    \label{fig:designed7}}
\end{center}
\end{figure}

\begin{figure}
\begin{center}
   \includegraphics[angle=0,width=0.4\textwidth]{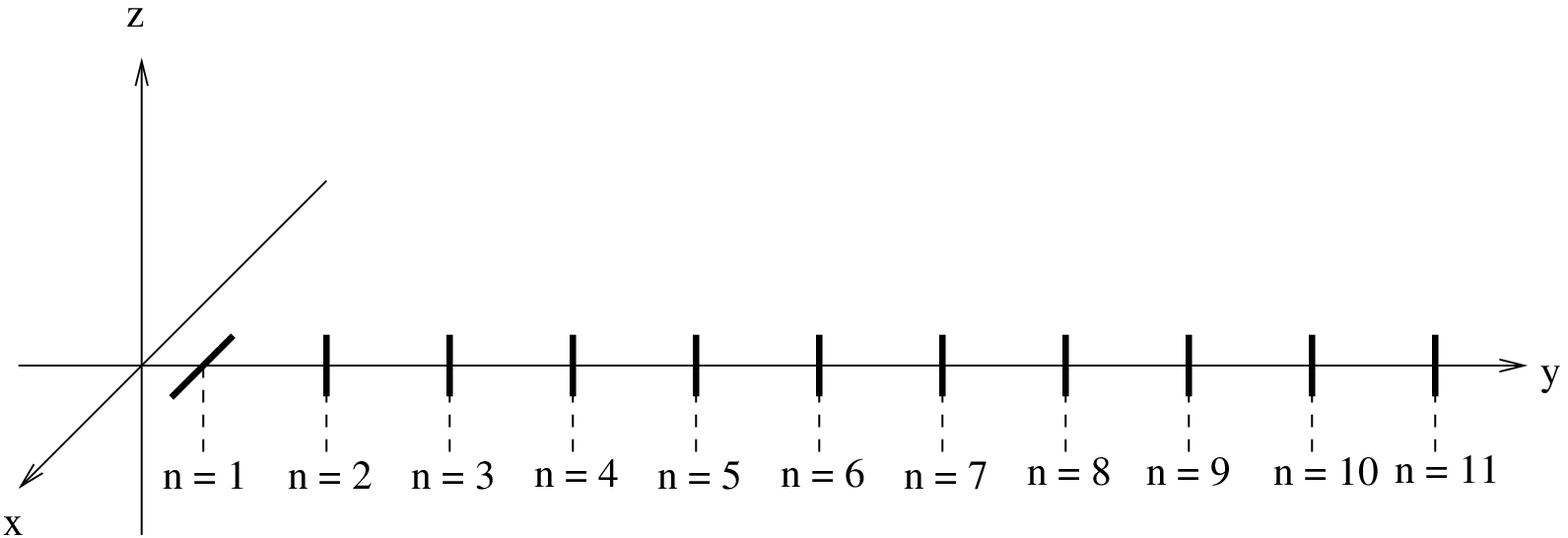}
   \caption{Dipole orientations for the off-broadside ($\theta_{ML}=70^{\circ}$ and $\phi_{ML}=-90^{\circ}$) example designed using BCS-IMDSM.
    \label{fig:designed8}}
\end{center}
\end{figure}

\begin{figure}
\begin{center}
   \includegraphics[angle=0,width=0.4\textwidth]{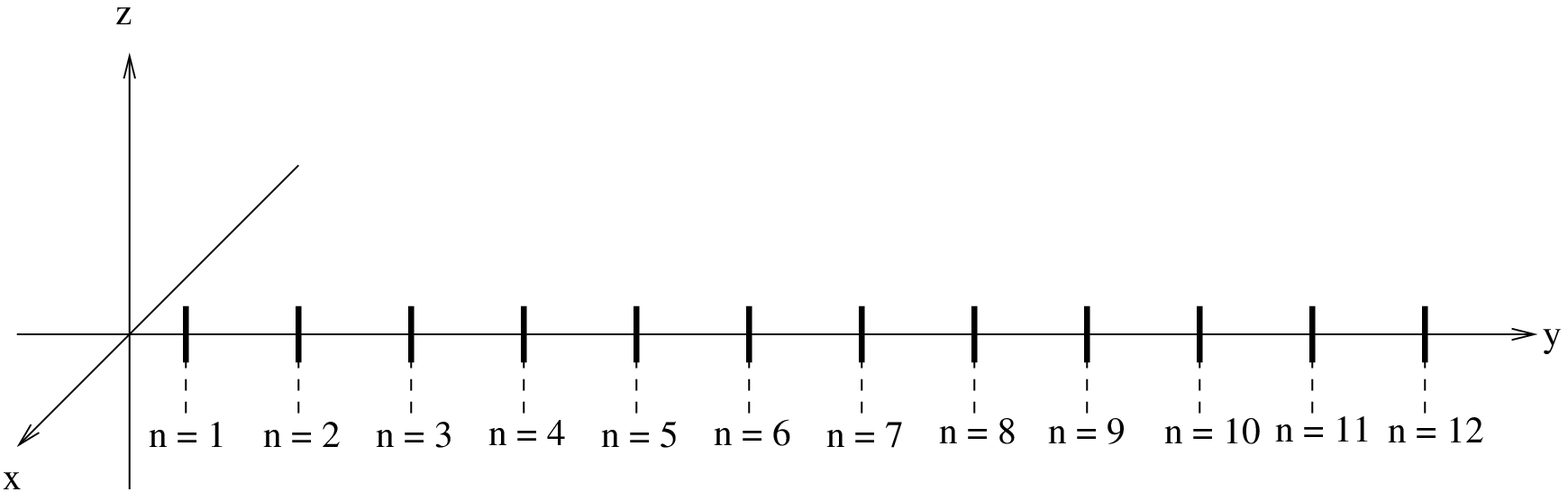}
   \caption{Dipole orientations for the off-broadside ($\theta_{ML}=70^{\circ}$ and $\phi_{ML}=-90^{\circ}$) example designed using AIRMS.
    \label{fig:designed9}}
\end{center}
\end{figure}

\begin{figure}
\begin{center}
   \includegraphics[angle=0,width=0.4\textwidth]{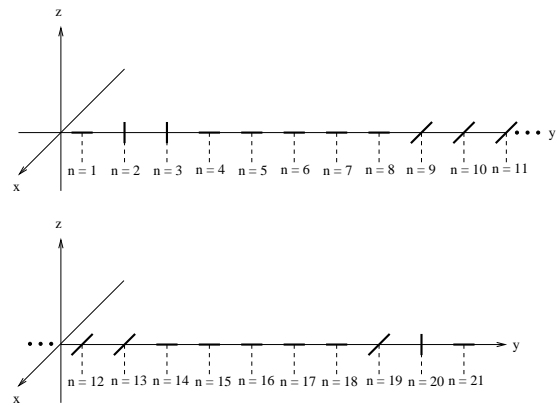}
   \caption{Dipole orientations for the off-broadside ($\theta_{ML}=70^{\circ}$ and $\phi_{ML}=-90^{\circ}$) ULA example.
    \label{fig:designedula3}}
\end{center}
\end{figure}

Firstly, the mainlobe for the AIRMS example is within $1^{\circ}$ of what was desired with the other three mainlobes being within $3^{\circ}$.  As a result, it is clear that the AIRMS has achieved a mainlobe direction closer to the desired direction than the comparison ULA.  Although the mainlobe for the CS-IMDSM and BCS-IMDSM are not as accurately located, they are still close enough ensuring that there is not significant suppression of signals from the desired location.  In addition, they are no worse than the comparison ULA in this regard.  This, along with the fact that sufficient sidelobe attenuation has been achieved, suggests that an acceptable response has been achieved by the proposed methods.  However, comparing the error values shows that the CS-IMSDM and BCS-IMSDM have given an approximation of the ideal achieved response that is worse than with the ULA.  Although, this is done using less dipoles (48$\%$ less for both methods) and a larger adjacent dipole separation.  We can also see that the AIRMS has given a similar reduction in the number of dipoles required (43$\%$ rather than 48$\%$), while also giving a better approximation of the ideal response than achieved by the ULA.  In this instance the BCS-IMDSM has proven to have the best computational efficiency.  For completeness the dipole locations and orientations are given in Tables \ref{tb:CSoffbroad2}-\ref{tb:reweightedoffbroad2} and Figures \ref{fig:designed7}-\ref{fig:designedula3}, respectively.  As for the previous examples the true dipole locations should be taken from the tables provided.


\begin{table}
\caption{\rm Dipole locations and orientations for the off-broadside ($\theta_{ML}=70^{\circ}$ and $\phi_{ML}=-90^{\circ}$) CS-IMDSM design example.} \centering
\begin{tabular}{|c|c|c|c|c|c|c|c|}\hline
n  & $d_{n}/\lambda$ &  n & $d_{n}/\lambda$ &  n & $d_{n}/\lambda$ & n & $d_{n}/\lambda$\\
\hline
1  &  0.19  &  4 &  3.09    &  7 & 6.03   & 10 & 9.05 \\
\hline
2  & 1.21 &  5  &   3.92   &  8 &  7.03  & 11 & 9.88 \\
\hline
3  & 2.13 &   6  &  5.00   & 9 & 8.03 \\
\cline{1-6}
\end{tabular}
\label{tb:CSoffbroad2}
\end{table}

\begin{table}
\caption{\rm Dipole locations and orientations for the off-broadside ($\theta_{ML}=70^{\circ}$ and $\phi_{ML}=-90^{\circ}$) BCS-IMDSM design example.} \centering
\begin{tabular}{|c|c|c|c|c|c|c|c|}\hline
n  & $d_{n}/\lambda$ &  n & $d_{n}/\lambda$ &  n & $d_{n}/\lambda$ & n & $d_{n}/\lambda$\\
\hline
1  &  0.38  &  4 &  3.44    &  7 & 6.27   & 10 & 8.88 \\
\hline
2  & 1.27 &  5  &   4.41   &  8 &  7.16  & 11 & 9.68 \\
\hline
3  & 2.44 &   6  &  5.36  & 9 & 8.02 \\
\cline{1-6}
\end{tabular}
\label{tb:CSoffbroad21}
\end{table}

\begin{table}
\caption{\rm Dipole locations and orientations for the off-broadside ($\theta_{ML}=70^{\circ}$ and $\phi_{ML}=-90^{\circ}$) AIRMS design example.} \centering
\begin{tabular}{|c|c|c|c|c|c|c|c|}\hline
n  & $d_{n}/\lambda$ &  n & $d_{n}/\lambda$ &  n & $d_{n}/\lambda$ & n & $d_{n}/\lambda$\\
\hline
1  &  0  &  4 &  2.63    &  7 & 5.40   & 10 & 8.13 \\
\hline
2  & 0.90 &  5  &   3.50   &  8 &  6.47  & 11 & 9.07 \\
\hline
3  & 1.83 &   6  &  4.60   & 9 & 7.33  & 12 & 10\\
\hline
\end{tabular}
\label{tb:reweightedoffbroad2}
\end{table}

\subsection{Discussion}
This subsection presents a discussion of the main results in light of the implications for optimal parameter selection.  These points can be summarised as follows:
\begin{enumerate}
  \item From the broadside design example it can be seen that increasing the value of $M$ always increases
  the computation time as more grid points are being considered.  For CS or BCS it would be reasonable to
  expect that increasing $M$ would improve the solution in terms of sparsity and desirability of the achieved
  response.  The iterative nature of the algorithms makes it harder to predict the effects on error
  between the reference and achieved responses and the number of dipoles required.  Experience suggests that
  $M=301$ is the best tradeoff to make.
  \item The iterative nature of the algorithms has also made it difficult to predict the effects of varying
  the value of $\alpha$ for the CS-IMDSM and AIRMS.  It is worth noting two points.  Firstly a
  value of $\alpha=0$ would mean that the approximation of the reference pattern would have to be exact.
  This is unlikely to be possible when the ideal response is used.  Secondly, a value of $\alpha=1$ will
  result in a response of all zeros and no dipoles being used, as $||[0, 0, ..., 1, ..., 0, 0]^{T}~-~[0, 0, ..., 0, ..., 0, 0]^{T}||_{2}=1$.
  \item The value of $L$ has to be large enough to consider all the angular and polarisation points of
  interest, as an acceptable response can not be guaranteed for the points not directly considered.
  Increasing $L$ further when this has been achieved adds computational complexity for no further
  gain in desirability of the array's response.
  \item The off broadside design examples provided indicate that one of the three methods (or alternative the method for designing a comparison uniform SST array) can be used for all off-broadside mainlobe directions of interest.  However, a single method can not be guaranteed to perform best in all cases.
\end{enumerate}

It is also worth considering the problem of selecting which of the three proposed methods should be used in a given situation.  There are 4 criteria to be considered: guarantees of a solution, the sparsity introduced, error between the reference and designed responses and the computational efficiency.  These 4 points are now considered in turn and recommendations made about which method to use.
\begin{enumerate}
  \item Guarantees of a solution:  The results provided show that the AIRMS was the only one not to always give a solution.  This would suggests using one of the other two methods when guarantees of a solution is the overriding factor.  The selection of which of the remaining two methods should be used depends on which of the remaining criteria are prioritised.
  \item The sparsity introduced: The results given above indicate that the CS-IMDSM tends to give the sparsest solution (followed by the BCS-IMDSM and then AIRMS) so should be selected when this criterion is the most important.
  \item Error between reference and designed responses: In terms of the amount of error between the reference and designed responses the BCS-IMDSM has been shown to give the best performance. This can be explained by the fact that the less dipoles used the more error is expected and the BCS-IMDSM method had lower levels of sparsity than the CS-IMDSM (while the AIRMS was not always guaranteed to give a solution).
  \item The computational efficiency:  If computational efficiency is prioritised over the other criteria the authors would suggest considering the BCS-IMDSM as the results consistently show it is the most efficient method (followed by AIRMS, when it gives a solution, and CS-IMDSM).
\end{enumerate}

\section{Conclusions}\label{sec:con}
In this work the problem of designing sparse SSSTA has been addressed for the first time.
Novel CS and BCS based approaches have been proposed to solve the problem of simultaneously optimising
dipole locations and orientations,
with a minimum spacing being used to avoid co-located dipoles.  Design examples have been provided and
show that an accurate approximation of a reference pattern can be achieved using fewer dipoles than a
comparable uniform SST array (38$\%$-86$\%$ reduction in the number of dipoles). This work has focused
on the design of linear SSSTAs for a single signal polarisation of interest.  In order to fully control
a wide range of signal polarisations a planar array may be necessary.  Extending the proposed approaches to
this case is seen as an area for future research.

\section*{Appendix}
\subsection{Single Task Bayesian Compressive Sensing for Spatially Stretched Sparse Tripole Array Design}
When looking to use ST-BCS to design SSSTA the problem can be considered in a similar form to what is done when designing traditional sparse arrays \cite{Oliveri11a,Oliveri12,Viani13,Oliveri12a}:
\begin{equation}\label{eq:bcs1}
  \textbf{p}_{r}-\textbf{Sw} = \textbf{D}
\end{equation}
where $\textbf{D}$ is a zero mean Gaussian error vector.  The variance of $\textbf{D}$ is proportional to the limit placed on the allowed error between the desired and achieved response, i.e. $\sigma^{2}\propto\alpha$.  These complex values can be split into real and imaginary parts giving
\begin{equation}\label{eq:bcs2}
  \hat{\textbf{p}}_{r}-\tilde{\textbf{S}}\tilde{\textbf{w}} = \tilde{\textbf{D}}
\end{equation}
where
\begin{equation}\label{eq:bcs4}
  \tilde{\textbf{w}} = [R(\textbf{w}),-I(\textbf{w})]^{T}
\end{equation}
and
\begin{equation}\label{eq:bcs5}
  \tilde{\textbf{D}}=[R(\textbf{D}),I(\textbf{D})]^{T}.
\end{equation}
Now model $\hat{\textbf{p}}_{r}$ as a Gaussian likelihood

\begin{equation}\label{eq:bcs6}
  \mathcal{P}(\hat{\textbf{p}}_{r}|\tilde{\textbf{w}},\sigma^{2})=\frac{1}{(2\pi\sigma^{2})^{\frac{L}{2}}}e^{-\frac{1}{2\sigma^{2}}||\hat{\textbf{p}}_{r}-\tilde{\textbf{S}}\tilde{\textbf{w}}||^{2}_{2}}.
\end{equation}

The problem of finding the optimal sensor locations is then solved by maximising the a-posteriori probability $\mathcal{P}(\tilde{\textbf{w}},\sigma^{2}|\hat{\textbf{p}}_{r})$ while also enforcing a belief that the weight coefficient vector should also be sparse.

This sparse belief can be enforced by using the Gaussian hierarchial prior
\begin{equation}\label{eq:bcs7}
  \mathcal{P}(\tilde{\textbf{w}}|\tilde{\textbf{a}})=(2\pi)^{-3M}\prod_{m=1}^{6M}\sqrt{\tilde{a}_{m}}\Bigg(-\frac{\tilde{a}_{m}\tilde{w}_{m}^{2}}{2}\Bigg),
\end{equation}
where $\tilde{a}_{m}$ is the hyperparameter that determines whether $\tilde{w}_{m}$ is zero-valued or not. To be able to fully evaluate (\ref{eq:bcs7}) further definitions have to be made, i.e. the hyperpriors over $\tilde{\textbf{a}}$ and $\sigma^{2}$, where $\tilde{\textbf{a}}=[\tilde{a}_{1},\tilde{a}_{2}, \ldots, \tilde{a}_{6M}]^{T}$.  Note, there are $6M$ hyperparameters as the real and imaginary parts of the weight coefficients for the $3M$ dipoles are being considered separately.  For the same reason the limit in \eqref{eq:bcs7} is also $6M$.  The hyperparameters are given by the following Gamma distributions:
\begin{equation}\label{eq:bcs8}
  \mathcal{P}(\tilde{\textbf{a}})=\prod_{m=1}^{6M}G(\tilde{a}_{m}|\beta_{ST-3},\beta_{ST-4})
\end{equation}
and
\begin{equation}\label{eq:bcs9}
  \mathcal{P}(\sigma^{2})=G(\sigma^{-2}|\beta_{ST-5},\beta_{ST-6}).
\end{equation}

The solution to the problem of maximising $\mathcal{P}(\tilde{\textbf{w}},\sigma^{2}|\hat{\textbf{p}}_{r})$ can now be found by following the methodology detailed for the RVM \cite{Ji08,Tipping01,Tipping03}, which will be briefly summarised below.

It is known that the posterior can be written as
\begin{equation}\label{eq:bcs10}
 \mathcal{P}(\tilde{\textbf{w}},\tilde{\textbf{a}},\sigma^{2}|\hat{\textbf{p}}_{r})=\mathcal{P}(\tilde{\textbf{w}}|\hat{\textbf{p}}_{r},\tilde{\textbf{a}},\sigma^{2})\mathcal{P}(\tilde{\textbf{a}},\sigma^{2}|\hat{\textbf{p}}_{r})
\end{equation}
and from (\ref{eq:bcs6}) and (\ref{eq:bcs7})
\begin{eqnarray}\label{eq:bcs11}\nonumber
  \mathcal{P}(\tilde{\textbf{w}}|\hat{\textbf{p}}_{r},\tilde{\textbf{a}},\sigma^{2}) = \frac{\mathcal{P}(\hat{\textbf{p}}_{r}|\tilde{\textbf{w}},\sigma^{2})\mathcal{P}(\tilde{\textbf{w}}|\tilde{\textbf{a}})}{\mathcal{P}(\hat{\textbf{p}}_{r}|\tilde{\textbf{a}},\sigma^{2})}\\
 = \frac{1}{(2\pi)^{3M+\frac{1}{2}}\sqrt{\det(\boldsymbol\Sigma)}}e^{-\frac{(\tilde{\textbf{w}}-\boldsymbol\mu)^{H}\boldsymbol\Sigma^{-1}(\tilde{\textbf{w}}-\boldsymbol\mu)}{2}}.
\end{eqnarray}
The posterior mean and variance are given respectively by
\begin{equation}\label{eq:bcs12}
  \boldsymbol\mu=\frac{\boldsymbol\Sigma\tilde{\textbf{S}}^{T}\hat{\textbf{p}}_{r}}{\sigma^{2}}
\end{equation}
and
\begin{equation}\label{eq:bcs13}
  \boldsymbol\Sigma=((\tilde{\textbf{S}}^{T}\tilde{\textbf{S}}/\sigma^{2})+\textbf{A})^{-1},
\end{equation}
where $\textbf{A}=\text{diag}(\tilde{a}_{1}, \tilde{a}_{2}, \ldots, \tilde{a}_{6M})$ is the diagonal matrix of the $6M$ hyperparameters.

A delta function at the values of $\tilde{\textbf{a}}$ and $\sigma^{2}$ that maximise $\mathcal{P}(\tilde{\textbf{a}},\sigma^{2}|\hat{\textbf{p}}_{r})$ can be used to approximate $\mathcal{P}(\tilde{\textbf{a}},\sigma^{2}|\hat{\textbf{p}}_{r})$ (i.e. a point estimate of $\mathcal{P}(\tilde{\textbf{a}},\sigma^{2}|\hat{\textbf{p}}_{r})$ for the most probable values of $\tilde{\textbf{a}}$ and $\sigma^{2}$).  It is also known that
\begin{equation}\label{eq:bcs14}
  \mathcal{P}(\tilde{\textbf{a}},\sigma^{2}|\hat{\textbf{p}}_{r})\propto\mathcal{P}(\hat{\textbf{p}}_{r}|\tilde{\textbf{a}},\sigma^{2})\mathcal{P}(\tilde{\textbf{a}})\mathcal{P}(\sigma^{2}).
\end{equation}
With uniform values of $\beta_{ST-3}, \beta_{ST-4}, \beta_{ST-5}$ and $\beta_{ST-6}$, $\mathcal{P}(\tilde{\textbf{a}})$ and $\mathcal{P}(\sigma^{2})$ become constant.  Therefore, maximising $\mathcal{P}(\tilde{\textbf{a}},\sigma^{2}|\hat{\textbf{p}}_{r})$ is equivalent to maximising $\mathcal{P}(\hat{\textbf{p}}_{r}|\tilde{\textbf{a}},\sigma^{2})$.  This can be solved by following a type II likelihood maximisation procedure to maximise the likelihood function given by
\begin{eqnarray}\label{eq:bcs15}\nonumber
  \mathcal{L}(\tilde{\textbf{a}},\sigma^{2})&=&\log[\mathcal{P}(\hat{\textbf{p}}_{r}|\tilde{\textbf{a}},\sigma^{2})]\\
&=&-\frac{1}{2}[6M\log2\pi+\log|\textbf{C}|+\hat{\textbf{p}}_{r}^{T}\textbf{C}^{-1}\hat{\textbf{p}}_{r}]
\end{eqnarray}
where
\begin{equation}\label{eq:bcs15a}
  \textbf{C}=\sigma^{2}\textbf{I}+\tilde{\textbf{S}}\textbf{A}^{-1}\tilde{\textbf{S}}^{T}.
\end{equation}
This allows the optimal values $\tilde{\textbf{a}}_{opt}$ and $\sigma^{2}_{opt}$ to be obtained.

The $m^{th}$ optimal weight coefficients are then given by
\begin{eqnarray}\label{eq:bcs16}
  \tilde{w}_{opt,x,m} &=& \tilde{w}_{opt,m} + j\tilde{w}_{opt,m+m}, \\ \label{eq:bcs17}
  \tilde{w}_{opt,y,m} &=& \tilde{w}_{opt,m+2m} + j\tilde{w}_{opt,m+3m}, \\ \label{eq:bcs18}
  \tilde{w}_{opt,z,m} &=& \tilde{w}_{opt,m+4m} + j\tilde{w}_{opt,m+5m},
\end{eqnarray}

where $\tilde{w}_{opt,m}$ is the $m^{th}$ entry of
\begin{equation}\label{eq:bcs19}
  \tilde{\textbf{w}}_{opt} = \frac{1}{\sigma^{2}_{opt}}\Bigg(\frac{\tilde{\textbf{S}}^{T}\tilde{\textbf{S}}}{\sigma^{2}_{opt}}+\textbf{A}_{opt}\Bigg)^{-1}\tilde{\textbf{S}}^{T}\hat{\textbf{p}}_{r}.
\end{equation}

\section*{Acknowledgments}

\noindent We appreciate the support of the UK Engineering and Physical Sciences Research Council (EPSRC) via the project Bayesian Tracking and Reasoning over Time (BTaRoT) grant EP/K021516/1.  We would like to thank the associate editor for handling our paper and acknowledge the anonymous reviewers' comments that have helped improve this work.


\newpage
\begin{IEEEbiography}
[{\includegraphics[width=1.29in,height=1.39in,trim = 15mm 14mm -10mm
0mm]{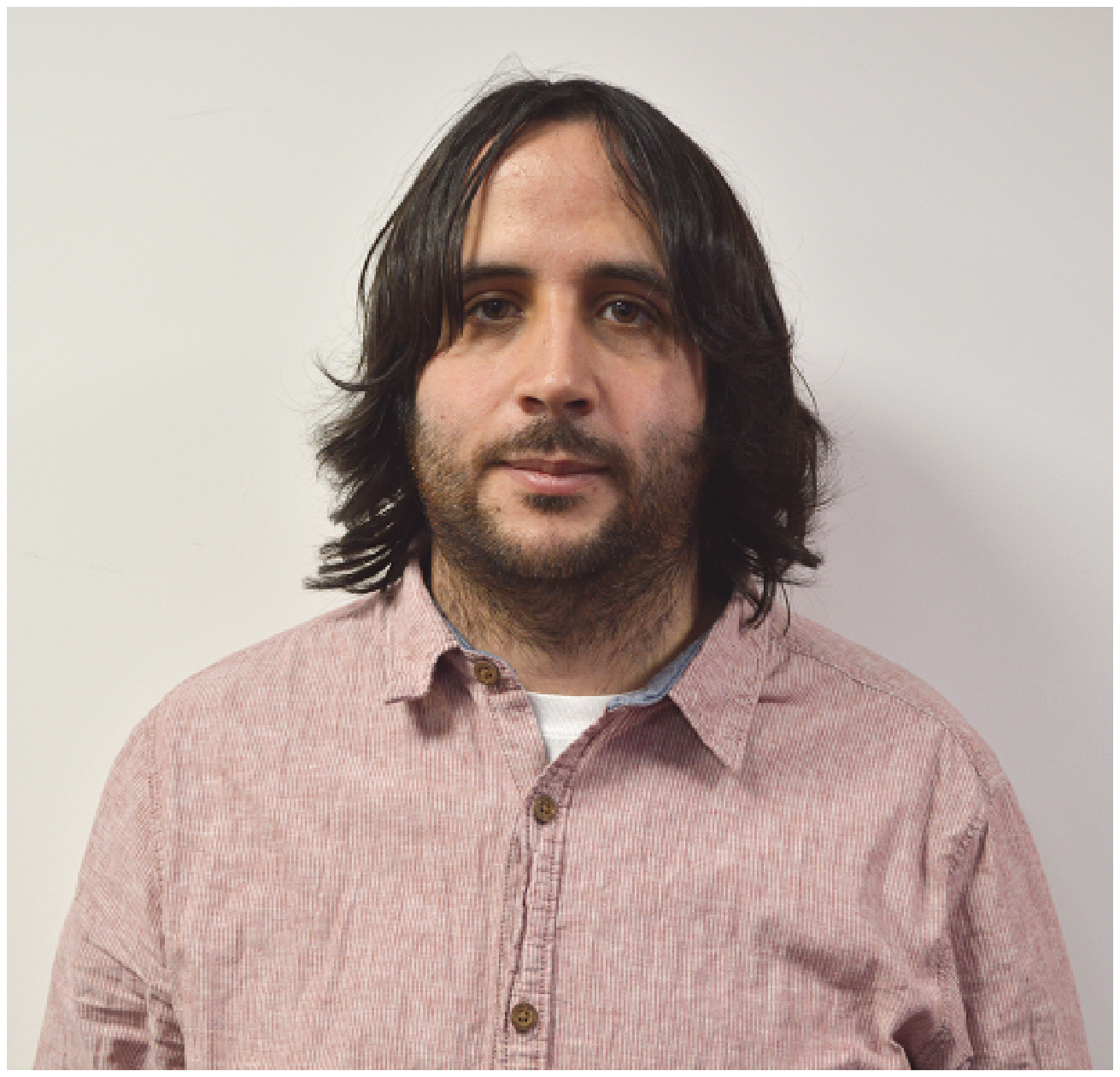}}] {Matthew Hawes} received his MEng and
PhD degree from the Department of Electronic and Electrical
Engineering at the University of Sheffield in 2010 and 2014,
respectively.  Since then he has been employed as a research
associate in the Department of Automatic Control and Systems
Engineering at the same university.  He is currently working on the
EU funded SETA project, the main scope of which is the development of
models, methods and a platform for mobility prediction, congestion
avoidance and sensor data fusion for smart cities. His research
interests include array signal processing, localisation and
tracking, big data, modelling complex systems, intelligent
transportation systems, mobility, data fusion, sequential Monte
Carlo methods and Markov chain Monte Carlo methods.
\end{IEEEbiography}
\vspace*{-3cm}
\begin{IEEEbiography}
[{\includegraphics[width=1.43in,height=1.39in,trim = 15mm 14mm -10mm
0mm]{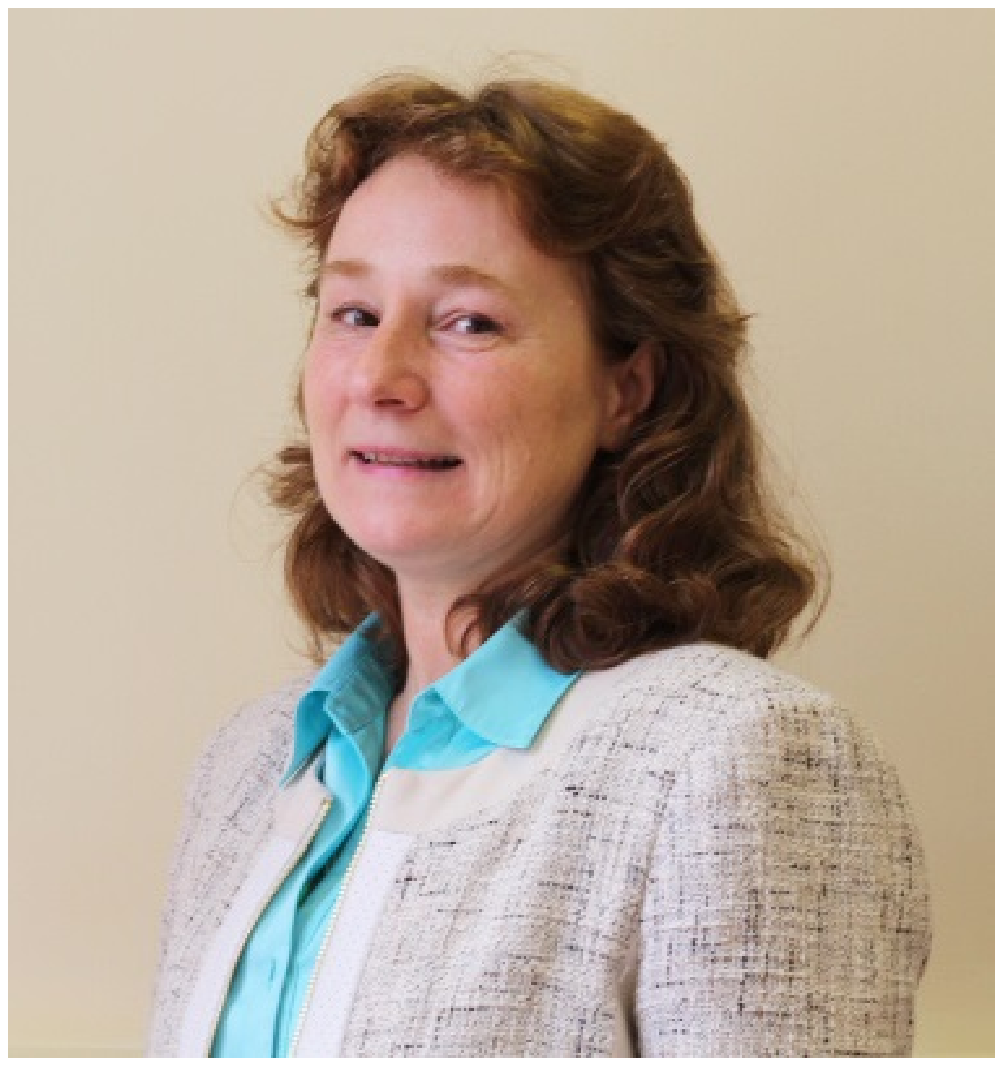}}] {Lyudmila Mihaylova} (M'98,
SM'2008) is Professor of Signal Processing and Control at the
Department of Automatic Control and Systems Engineering at the
University of Sheffield, United Kingdom. Her research is in the
areas of machine learning and autonomous systems with various
applications such as navigation, surveillance and sensor network
systems. She has given a number of talks and tutorials, including
the plenary talk for the IEEE Sensor Data Fusion 2015 (Germany),
invited talks University of California, Los Angeles, IPAMI Traffic
Workshop 2016 (USA), IET ICWMMN 2013 in Beijing, China. Dr.
Mihaylova is an Associate Editor of the IEEE Transactions on
Aerospace and Electronic Systems and of the Elsevier Signal
Processing Journal. She was elected in March 2016 as a president of
the International Society of Information Fusion (ISIF). She is on
the board of Directors of ISIF and a Senior IEEE member. She was the
general co-chair IET Data Fusion $\&$ Target Tracking 2014 and 2012
Conferences, Program co-chair for the 19th International Conference
on Information Fusion, Heidelberg, Germany, 2016, academic chair of
Fusion 2010 conference.
\end{IEEEbiography}
\vspace{-3cm}
\begin{IEEEbiography}
[{\includegraphics[width=1.3in,height=1.39in,trim = 15mm 14mm -10mm
0mm]{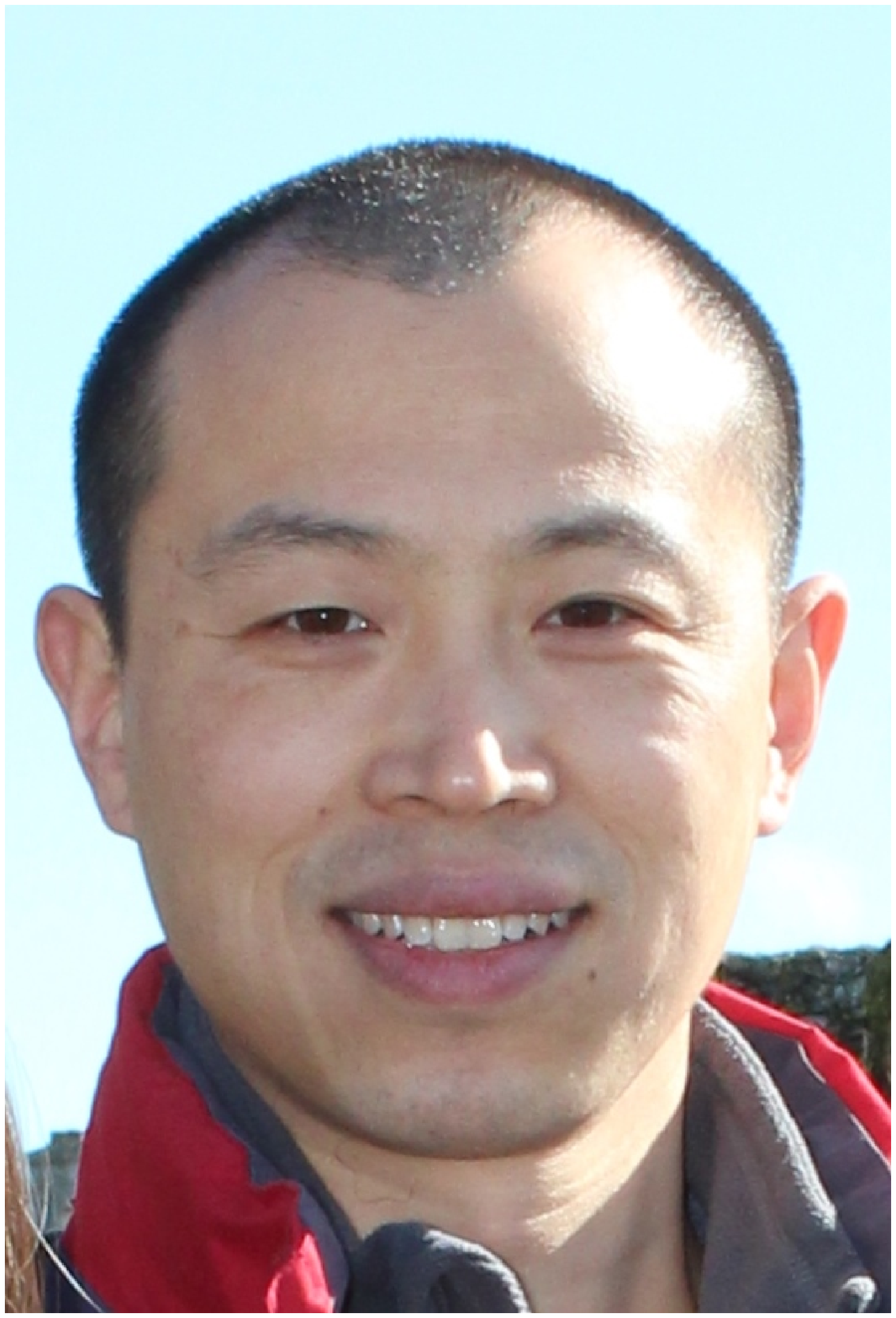}}]
{Wei Liu} Wei Liu received his B.Sc. in 1996 and L.L.B. in 1997, both from Peking University, China, M.Phil. from University of Hong Kong, in 2001, and Ph.D. in 2003 from the School of Electronics and Computer Science, University of Southampton, UK. He later worked as a postdoc at Imperial College London. Since September 2005, he has been with the Department of Electronic and Electrical Engineering, University of Sheffield, UK, first as a lecturer, and then a senior lecturer. His research interests are in sensor array signal processing, blind signal processing, multivariate signal processing and their various applications in wireless communications, radar, sonar, satellite navigation, human computer interface and renewable energy exploitation, etc. He has now published more than 200 journal and conference papers, three book chapters, and a research monograph about wideband beamforming (``Wideband Beamforming: Concepts and Techniques'', John Wiley $\&$ Sons, March 2010). He is a senior member of IEEE, an elected member of the Digital Signal Processing Technical Committee of the IEEE Circuits and Systems Society and the Sensor Array and Multichannel Signal Processing Technical Committee of the IEEE Signal Processing Society, currently an associate editor for IEEE Trans. Signal Processing and IEEE Access and an editorial board member of the journal Frontiers of Information Technology $\&$ Electronic Engineering .
\end{IEEEbiography}


\end{document}